# Two-dimensional bimetal-embedded expanded phthalocyanine monolayers: a class of multifunctional materials with fascinating properties

De-Bing Long,[1] Nikolay V. Tkachenko,[2] Qingqing Feng,[3] Xingxing Li,[3] Alexander I. Boldyrev,[2] Jinlong Yang,[3] and Li-Ming Yang*[1]

[1]*Key Laboratory of Material Chemistry for Energy Conversion and Storage, Ministry of Education; Hubei Key Laboratory of Materials Chemistry and Service Failure; Hubei Key Laboratory of Bioinorganic Chemistry and Materia Medica; Hubei Engineering Research Center for Biomaterials and Medical Protective Materials; School of Chemistry and Chemical Engineering, Huazhong University of Science and Technology, Wuhan, 430074, China. (email: Lmyang@hust.edu.cn);* [2]*Department of Chemistry and Biochemistry, Utah State University, Logan, Utah 84322, USA;* [3]*Department of Chemical Physics, Hefei National Laboratory for Physical Sciences at the Microscale, and Synergetic Innovation Center of Quantum Information & Quantum Physics, University of Science and Technology of China, Hefei, Anhui 230026, China*

**Abstract**: The expanded phthalocyanine (EPc) single-layer sheets with double transition metals (labeled as $TM_2EPc$, TM = Sc–Zn) are predicted to be a new class of two-dimensional (2D) metal-organic materials with a series of favorable functional properties by means of systematic first-principle calculations and molecular dynamics simulations. The strong coordination between metal and EPc substrate accounts for the excellent structural stability. Chemical bonding analysis has demonstrated the absence of TM-TM bonding. Each metal center is isolated, but connected to the organic framework by four 2c-2e TM-N σ-bonds to form an extended 2D network. Unexpectedly, it is found that the $V_2EPc$ is an antiferromagnetic metal with Dirac cone, while $Cr_2EPc$ exhibits ferromagnetic Dirac half-metallicity, which is not common in 2D materials. Excitingly, the ferromagnetic $Cr_2EPc$ and antiferromagnetic $Mn_2$– and $Fe_2$–EPc have high magnetic transition temperatures of 223, 217, and 325 K, respectively, which are crucial for the practical applications of spintronics. $Cr_2EPc$ can maintain the Dirac half-metallicity under −6 % ~ 2 % biaxial strains, and $Fe_2EPc$ can transform from semiconductor to half-metal by applying −6 % ~ −10 % compressive strains. Additionally, the $TM_2EPc$ monolayers exhibit a full response to visible light and some materials have strong absorption in the ultraviolet and infrared regions in addition to visible light, showing extraordinary solar light-harvesting ability. Notably, the designed type-II heterojunctions $Fe_2EPc/SnC$, $Co_2EPc/GeS$, and $Ni_2EPc/2H-WSe_2$ have high power conversion efficiency (PCE > 15%), especially the PCE of $Ni_2EPc/2H-WSe_2$ reaches 25.19%, which has great potential in solar cell applications. All these desired properties render 2D $TM_2EPc$ monolayers promising candidates for future applications in nanoelectronics, spintronics, optoelectronics, and photovoltaic devices.

# 1. Introduction

Since the discovery of graphene, two-dimensional (2D) materials have caused great innovations in the fields of physics, chemistry, materials, nanotechnology, biology and energy due to their unique structural characteristics, excellent properties and promising applications in nanoelectronics, catalytic reaction, energy storage and conversion, sensors and biomedicine, etc.[1] However, many 2D materials are intrinsically nonmagnetic in their pristine forms, which limits their further applications in spintronic devices. Embedding transition metal (TM) atoms with unfilled $d$ shells may introduce magnetism into these 2D systems,[2-3] but it is difficult to control the distributions precisely, especially the TM atoms tend to cluster due to strong $d$–$d$ interactions. For these reasons, researchers are actively looking for new monolayers which TM atoms can be distributed in their backbone regularly and separately to exhibit intrinsically magnetism. Phthalocyanine (Pc) is an ideal template allowing the flexibility of embedding TM atoms in its pores, and the TM atoms are located at center of the host Pc and coordinated with four nitrogen atoms, so they are prevented from clustering when TMPc molecules form 2D structures. The 2D TMPc monolayers have recently attracted considerable attention due to their well-ordered geometries, flexibility in synthesis, and promising applications in gas capture and storage, highly active catalysts, optoelectronic and spintronic devices.[4-10] The successful synthesis of the single-layer FePc,[4] CuPc,[11] and MnPc[12] monolayers undoubtedly paves the way for exploring 2D TMPc monolayers. Furthermore, the properties of 2D TMPc monolayers are closely related to the central metal species, and thus the electronic and magnetic properties as well as the catalysis and adsorption can be modulated through substitution of diverse metal atoms.[9] This indicates the feasibility of designing and synthesizing function-oriented 2D TMPc monolayers, which will further expand their potential applications.

Recently, a rectangular-shaped expanded Pc congener containing Mo or W bimetal centers and four isoindoles ring moieties was synthesized experimentally.[13] Inspired by the novel TM$_2$EPc molecule structure, we want to know whether TM dimers can be anchored on the expanded Pc to form stable 2D TM$_2$EPc single-layer monolayers, and if so, what physicochemical properties will such 2D bimetal expanded Pc monolayers have? Based on this idea, we carried out systematically theoretical investigation on the physicochemical properties of the 2D TM$_2$EPc (TM = Sc–Zn) monolayers. Such 2D TM$_2$EPc monolayers exhibit unique stability, electronic, magnetic, and optical properties, and are expected to be synthesized and have practicable applications in the future.

Herein by employing first-principles calculations, molecular dynamics, and Monte Carlo simulations, we predicted the new two-dimensional materials: TM$_2$EPc monolayers, having high stabilities, unexpected and fascinating properties. The effects of elastic strain engineering on the magnetic, electronic, and optical properties are comprehensively and deeply studied. Nonnegative phonon spectra and simulated annealing demonstrated their lattice dynamical and thermal stabilities. Chemical bonding analysis shows that there are six special five-center two-electron (5c-2e) π-bonds in these structures. $V_2$EPc and $Cr_2$EPc monolayers are found to be antiferromagnetic Dirac metal and ferromagnetic Dirac half-metal, respectively. $Cr_2$–, $Mn_2$–, and $Fe_2$–EPc monolayers possess high magnetic transition temperatures, showing great possibility of the practical applications at or close to room temperature. Optical absorption coefficients and in-plane Young's modulus calculations indicate that the 2D $TM_2$EPc monolayers have considerable mechanical and optical anisotropy. A high PCE of 25.19% indicates that $Ni_2$EPc/2H-$WSe_2$ is a promising candidate for efficient components of solar energy harvesting and photoinduced charge-carrier generation in photovoltaic cells.

## 2. Computational methodologies

All first-principles calculations based on spin-polarized density functional theory (DFT) were performed using the projector augmented wave (PAW) method[14] as implemented in Vienna *Ab initio* Simulation Package (VASP).[15] The generalized gradient approximation (GGA) of Perdew–Burke–Ernzerhof (PBE)[16] was used for the exchange-correlation functional to treat the interactions between electrons. To properly treat the strong-correction effects of *d* orbits in TM atoms, a PBE+U strategy[17] with the value of 4 eV for correlation U and 1 eV for exchange energy J was adopted to describe the Coulomb and exchange corrections, which has been used in study of similar TM-organic complex system.[9] A 500 eV cutoff for the plane wave expansion was adopted in all the computations. Geometries were optimized until the convergence criteria of energy and force were less than $10^{-6}$ eV and 0.01 eV/Å, respectively. In order to calculate the magnetic coupling between two TM atoms of possible magnetic configurations, a rectangular supercell containing four TM atoms was adopted. The Γ-central Monkhorst–Pack[18] special *k*-point meshes of 5 × 5 × 1 and 3 × 4 × 1 were used for the primitive unit cell and rectangular supercell, respectively, to represent their reciprocal spaces. The van der Waals (vdW) interactions were also fully considered and corrected via Grimme's semiempirical dispersion corrections (DFT-D3).[19] The periodic boundary condition was applied to simulate the 2D $TM_2$EPc monolayers, and a vacuum space of more than 15 Å along *z* direction was used to prevent interactions between neighboring slabs.

The phonon calculations of 2D $TM_2$EPc monolayers were performed by using the density functional perturbation theory (DFPT)[20] as implemented in PHONOPY package[21] interfaced to

VASP. To obtain reliable phonon dispersion results, further geometry optimization was performed with more stringent energy and force convergence criteria, especially the forces need to converge very well. A higher accuracy of $10^{-8}$ eV for energy convergence and $10^{-7}$ eV/Å for force convergence was applied during the phonon calculations. *Ab initio* molecular dynamics (AIMD) simulation was carried out at a series of different temperatures to assess the thermal stability of 2D $TM_2EPc$ monolayers. At each temperature, AIMD simulation in the constant number, volume, temperature (NVT) ensemble last for 10 ps with a time step of 1.0 fs, and the temperature was controlled by using the Nosé–Hoover thermostat method.[22]

In chemical bonding analysis, the $TM_2EPc$ molecular models were built and investigated firstly to get insight on chemical bonding of 2D $TM_2EPc$ monolayers. For $TM_2EPc$ molecular models, geometry optimization and frequency calculations were performed using Gaussian 16 software.[23] Optimized geometries are reported at the PBE0/LANL2DZ[24-27] level of theory. To understand the chemical bonding of investigated species, we carried out electron localization analysis at the same level of theory using the adaptive natural density partitioning (AdNDP) method[28] as implemented in the AdNDP 2.0 code.[29] Previously, AdNDP was shown to be insensitive to the level of theory or the basis set used. In addition, quantum theory of atoms in molecules (QTAIM)[30] analysis was done as implemented in Multiwfn code.[31] While for 2D $TM_2EPc$ monolayers, all solid-state calculations were carried out via VASP program. Calculations were performed solely for producing wave functions of the systems. The PBE+U ($U_{eff}$ = U–J, U = 4 eV, J = 1 eV) was applied for the electron correlation in the calculations. The Brillouin zone has been sampled by the Monkhorst-Pack method. Since the number of atoms in unit cell and supercell are different, various *k*-spaces were used. The 9 × 9 × 1 *k*-grid was used for calculating unit cell, while 3 × 4 × 1 *k*-grid was used for rectangular supercell containing four metal atoms. Energy cutoff was set to 500 eV, the partial band occupancies were set using Gaussian smearing with sigma parameter 0.1. To find the bonding pattern of the selected monolayers, the solid state adaptive natural density partitioning (SSAdNDP) algorithm[32] were implemented. The def2-TZVP atomic centered basis[33] was used for all elements to represent the projected plane wave density. The spillage parameter of occupied bands for this basis was no higher than 0.04%. The Visualization for Electronic and Structural Analysis software (VESTA, series 3)[34] was used for visualization of solid-state structures.

The band structure and density of states (DOS) were calculated at PBE+U level to characterize the electronic properties of 2D $TM_2EPc$. However, although the PBE+U can produce reliable lattice constants and correct magnetic ground states, it cannot fully compensate the underestimation of band gaps caused by the intrinsic PBE errors in the approximation of single-particle eigenvalues and the lack of derivative discontinuity.[35] Therefore, more accurate hybrid Heyd–Scuseria–Ernzerhof functional (HSE06)[36] was adopted to improve the reliability of the electronic band structure calculations. In the HSE06 method, a fraction of the exact screened Hartree–Fock (HF) exchange was incorporated into the PBE exchange using a mixing parameter $α = 0.25$.

For magnetic calculations, the rectangular supercell containing four TM atoms was applied to calculate the magnetic coupling between TM atoms. Four possible magnetic ordering configurations: ferromagnetic (FM), antiferromagnetic-1 (AFM1), antiferromagnetic-2 (AFM2) and antiferromagnetic-3 (AFM3) states were constructed, and the nearest, next-nearest and third-nearest exchange interactions were considered to compute the exchange energies and coupling parameters. The Monte Carlo (MC) simulation based on the classical Heisenberg model was used to calculate the Curie/Néel temperature ($T_C/T_N$) of phase transition in 2D TM$_2$EPc. In order to understand the origination of the magnetism in 2D TM$_2$EPc monolayers, the spatial distribution of the spin-polarized charge densities ($Δρ = ρ_↑ − ρ_↓$) were also analyzed.

The optical absorption coefficients of 2D TM$_2$EPc monolayers were obtained by calculating the frequency dependent real and imaginary parts of dielectric function. The number of bands was increased to 500 to guarantee enough empty bands for electronic excitation, and the complex shift $η$ in Kramers-Kronig transformation[37] was set to be 0.1 for the conversion between real and imaginary parts.

## 3. Results and discussion
### 3.1. Geometric structures and ground states of 2D TM$_2$EPc monolayers

The models of 2D TM$_2$EPc monolayers were constructed by employing the expanded phthalocyanine (marked by orange dotted line in Figure 1a and 1b) and phenyl (marked by purple dotted line in Figure 1a and 1b) as organic ligands, the 3$d$ transition metal series (Sc–Zn) as metal centers. The metal atoms are coordinated by two pyrrole nitrogen atoms (N1) and two amine nitrogen atoms (N2) of the macrocycle. Structural relaxation shows that such a framework can be classified as either Type-A or Type-B depending on the central metal atoms. Type-A is exactly planar without any wrinkle (Figure 1a), while Type-B, as shown in Figure 1b, four central atoms (two metal atoms and two N atoms) slightly protrude out of the plane, causing a buckled structure. Among the 2D TM$_2$EPc monolayers, five systems, namely, Cr$_2$–, Mn$_2$–, Fe$_2$–, Co$_2$–, and Ni$_2$–EPc, are belong to Type-A with space group *Cmmm* (No. 65), whereas the remaining five systems, Sc$_2$–, Ti$_2$–, V$_2$–, Cu$_2$– and Zn$_2$–EPc systems, are the structures of Type-B with space group *Cmm*2 (No. 35) (see Table S1).

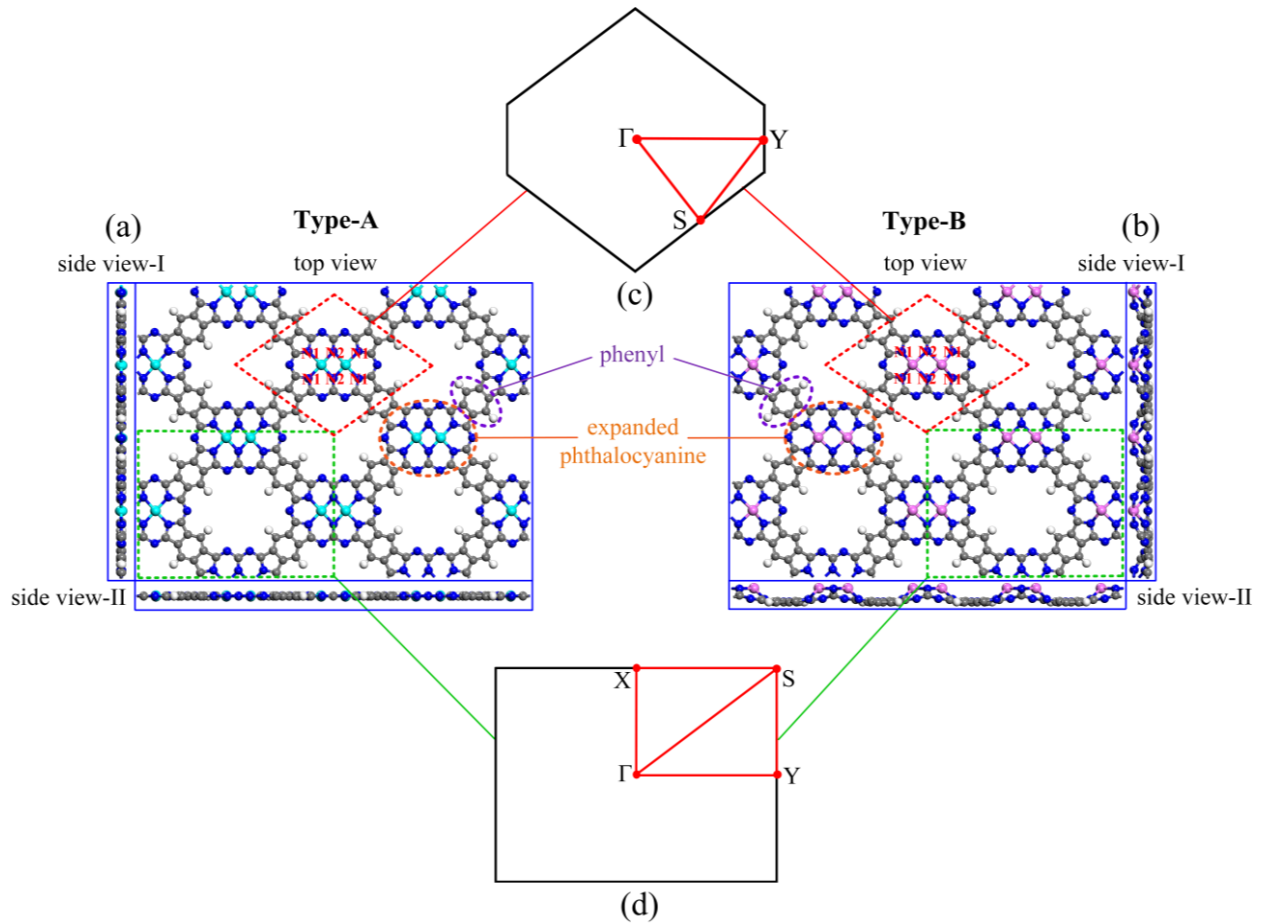

**Figure 1.** Top view and two different side view modes of the 2D TM$_2$EPc monolayers are displayed in the right, left and bottom panels in Figure 1a and 1b, respectively. N, blue; C, gray; H, white; the magenta balls in (a) and cyan balls in (b) are TM atoms. All the 2D TM$_2$EPc monolayers are divided into two categories: Type-A and Type-B. Type-A is planar structure, while Type-B is buckled. Among these 2D TM$_2$EPc monolayers, five systems, namely, Cr$_2$–, Mn$_2$–, Fe$_2$–, Co$_2$–, and Ni$_2$–EPc, are belong to Type-A, whereas the remaining five systems, Sc$_2$–, Ti$_2$–, V$_2$–, Cu$_2$–, and Zn$_2$–EPc, are the structures of Type-B. In the top view of (a) and (b), the primitive cell, rectangular supercell, organic linkers of the expanded phthalocyanine and phenyl are marked by red, green, orange, and purple dotted lines, respectively. (c), (d) The first Brillion zone in reciprocal space and high symmetry points (labeled by red dots) of primitive cell and rectangular supercell. For primitive cell, the high symmetric points are Γ (0, 0, 0), S (0.0, 0.5, 0.0), and Y (–0.5, 0.5, 0.0), while for supercell, the high symmetric points are Γ (0, 0, 0), X (–0.5, 0.0, 0.0), S (–0.5, 0.5, 0.0), and Y (0.0, 0.5, 0.0).

In our whole calculations, the rhombus primitive cell (marked by red dotted lines in Figure 1a) and rectangular supercell (marked by green dotted lines in Figure 1b) containing four metal atoms are adopted. Figure 1(c) and 1(d) are their corresponding first Brillion zone and high symmetric point paths in reciprocal space, respectively. Firstly, the primitive cells were used to distinguish the nonmagnetic (NM) and magnetic structures by performing spin-polarized DFT calculations. Then, the rectangular supercells were used to determine the magnetic ground states of the systems. The optimized results (see Table S1) show that $Sc_2$– and $Ni_2$– to $Zn_2$–EPc monolayers are nonmagnetic ground states; $V_2$– and $Cr_2$–EPc systems are AFM1 and FM orderings, respectively; whereas the magnetic ground states of remaining systems ($Ti_2$– and $Mn_2$– to $Co_2$–EPc) prefer to AFM3 configurations. Detailed description about the magnetic calculations will be present in Section 3.4. The lattice parameters a and b of the ground state structures are in the range of 19.67–19.97 Å and 15.20–15.34 Å, respectively. As shown in Figure S1, the magnetic moments exhibit the opposite trend respect to the distances between the central TM atoms ($d_{TM-TM}$ = 2.42–3.03 Å). The $Mn_2$EPc has the shortest distance between metal atoms (2.42 Å) and the largest magnetic moment (3.55 $\mu_B$). In addition, we can observe that the distances of TM–N1 ($d_{TM-N1}$ = 1.83–2.12 Å) are larger than that of TM–N2 ($d_{TM-N2}$ = 1.81–2.10 Å), and both of them show the same change rule as the single-bond radii of metal atoms.[39] After having determined the magnetic ground states, in the following sections, properties calculations are performed mainly based on these ground state structures.

## 3.2. Stability evaluation of 2D TM$_2$EPc monolayers

After exploring the geometric structures of 2D TM$_2$EPc monolayers, the ground states of each system were identified. In order to further explore the physicochemical properties of these ground state structures, the stability of 2D TM$_2$EPc monolayers are firstly evaluated from the following three aspects: lattice dynamic stability, thermal stability, and formation energies, cohesive energies, and binding energies.

### *3.2.1. Lattice dynamic stability*

The phonon dispersion curves along high symmetric points in the first Brillouin zone were calculated to verify the lattice dynamic stability of 2D TM$_2$EPc monolayers. As shown in Figure S2, there are no imaginary vibrational modes are observed in the entire Brillouin zone. This confirms their lattice dynamic stability. Furthermore, the highest phonon frequencies of TM$_2$EPc all exceed 3125 cm$^{-1}$, which are much higher than many reported 2D materials, such as MoS$_2$ (473 cm$^{-1}$),[40] graphene (1600 cm$^{-1}$),[41] Cu$_2$Si (420 cm$^{-1}$),[42] and AlB$_6$ (1150 cm$^{-1}$).[43] Remarkably, there is a considerably large gap between 1600 and 3125 cm$^{-1}$, which is extremely rare in 2D materials. For better understanding the contributions of each type of atoms to phonon vibration frequency, we display the phonon density of states (PhDOS) next to phonon dispersion spectra in

Figure S2. From PhDOS analysis, we can see that the contribution of PhDOS from C atoms is much larger than that from N, H, and metal atoms. A major and three minor PhDOS peaks are present near 750, 380, 1130, and 1388 cm$^{-1}$, respectively, corresponding to the dense regions in phonon spectra, which are mainly from the contributions of C and N atoms. The high frequency region (above 3125 cm$^{-1}$) is almost exclusively contributed by the H atoms, indicating the strong vibration of H; while the region below 1600 cm$^{-1}$ is mainly contributed by N, C, and metal atoms, followed by some contributions from the H atoms. The contribution of metal atoms locates in lower region (less than 700 cm$^{-1}$) due to their heavier mass. The results of phonon calculations indicate that 2D TM$_2$EPc monolayers are quite stable from the perspective of lattice dynamics.

*3.2.2. Thermal stability*

For practical applications of the proposed 2D TM$_2$EPc monolayers in nanoelectronics and spintronics at room temperature or even higher, we have performed the Born-Oppenheimer AIMD simulations in NVT ensemble to assess the thermal stability. In order to evaluate the highest temperature of each system can maintain structural integrity, a series of individual AIMD simulations with the duration of 10 ps were carried out at different temperatures. The fluctuations of total energy and temperature *versus* simulation time as well as the snapshots of these monolayers taken at the end of 10 ps are displayed in Figures S3–S12. We can see that the highest temperatures of each system can survive all exceed 1200 K, far above room temperature. The estimated melting points for 2D Sc$_2$– to Zn$_2$–EPc monolayers are 1300, 1800, 1600, 1400, 1500, 1400, 1200, 1800, 1400, and 1200 K, respectively. Below melting points, the total energies are slightly oscillated around their equilibrium positions and can maintain their structural integrity, once exceeding the melting points, the frameworks will appear significant distortion and start to melt. AIMD simulation results show that 2D TM$_2$EPc monolayers have high thermal stability, thus they hold the potential for practical applications as nanoelectronic and spintronic materials under ambient conditions and even higher temperature situation.

*3.2.3 Thermodynamic stability*

Apart from lattice dynamic and thermal stabilities, the formation energy $E_f$, cohesive energy $E_c$, and binding energy $E_b$ are also calculated to further evaluate the thermodynamic stability of these monolayers. The formation energy can reflect the difficulty of experimental preparation to a certain extent, and its negative value means that the preparation is an exothermic process. Figure S2 shows that the formation energies of the 2D TM$_2$EPc monolayers are in the range of –4.98 to 0.80 eV, and all the systems are negative except for Zn$_2$EPc monolayer (see Figure S13). These formation energies indicate the high stability and the possibility of experimental preparation of TM$_2$EPc monolayers. From Figure S13 we can see that the binding energies between TM and EPc are in the range of –5.49 ~ –10.86 eV, which is significantly smaller than the cohesive energies of the corresponding metal bulks (1.27–4.64 eV). The negative values of $E_b + E_c$ (–6.22 ~ –3.65 eV) indicates that the metal atoms can be stably embedded into the EPc substrate and are less likely to form clusters. This means that the 2D TM$_2$EPc monolayers have the potential as uniformly dispersed bimetal catalysts.

### 3.3. Chemical bonding analysis

To get deep insights on chemical bonding pattern of the 2D TM$_2$EPc monolayers, we firstly decided to investigate their molecular models. Herein, we took the Ni-containing structure for detailed analysis to illustrate chemical bonding of these molecular models. The Ni$_2$EPc molecule includes 232 valence electrons which can be localized into 116 two-electron bonding elements. The details of the process of model construction and optimization as well as the discussion about the AdNDP analysis of the Ni$_2$EPc molecule are shown in SI. By AdNDP analysis, we find that the chemical bonding pattern of the TM$_2$EPc molecules could be described in terms of classical Lewis bonding elements with four locally π-aromatic C$_6$-ring fragments and two π-conjugated C$_2$N$_3$ organic chains. From Table S4, we can see that the metal atom donates one electron to C$_2$N$_3$ chain, resulting the formation of three 5c-2e π-conjugated bonds, and the bonding between metal atoms and organic part occurs via eight 2c-2e TM-N σ-bonds. In addition, we have also performed the QTAIM analysis for all investigated molecular systems (see Figure S19). It is noteworthy that both methods have shown the absence of TM-TM bonding.

Following the clues obtained from molecular models, we analyzed periodic 2D TM$_2$EPc monolayers via SSAdNDP algorithm. The example of complete chemical bonding analysis of nonmagnetic 2D Ni$_2$EPc monolayer is shown in Figure 2. A bonding pattern similar to Ni$_2$EPc molecule is observed in Ni$_2$EPc monolayer. That is, in 2D Ni$_2$EPc monolayer, four $d$-type lone pairs on each Ni-atom (ON = 1.99–1.77 |e|) and six $s$-type lone pairs on N atoms (ONs = 1.87 |e|) are localized. The organic framework is bound via forty-six 2c-2e σ-bonds with ON = 1.96–1.94 |e| and six 2c-2e π-bonds with ON = 1.81–1.80 |e|. Each metal center is bounded to the organic framework by four 2c-2e TM-N σ-bonds with ON = 1.95–1.93 |e|. Also, we find that six delocalized 5c-2e π-bonds with ON = 1.97–1.64 |e| are responsible for conjugation in C$_2$N$_3$ fragment. The remaining electrons form six 6c-2e π-bonds which are responsible for aromaticity in C$_6$-ring. Almost the same bonding patterns are observed in other TM$_2$EPc monolayers. The difference between bonding is in the number of alpha and beta unpaired $d$-electrons of the metal atom. The results of SSAdNDP analysis for TM$_2$EPc monolayers are shown in Table S5. The main difference in bonding patterns is the number of *d*-type lone pairs on metal atom, which is also the fundamental origination resulting in 2D TM$_2$EPc monolayers exhibit different magnetic moments. Another significant difference is the bond length between nitrogen (that form a diamond with two metal atoms and opposite nitrogen) and the closest to it carbon atom. We find that for planar structures this bond is indeed shorter than the single C-N bond, but for buckled structures (except Zn$_2$EPc structure) this bond is slightly elongated. It can be explained by more negative charge on nitrogen atom while extra electrons can contribute to antibonding C-N orbits, resulting in longer bond distance (Table S3).

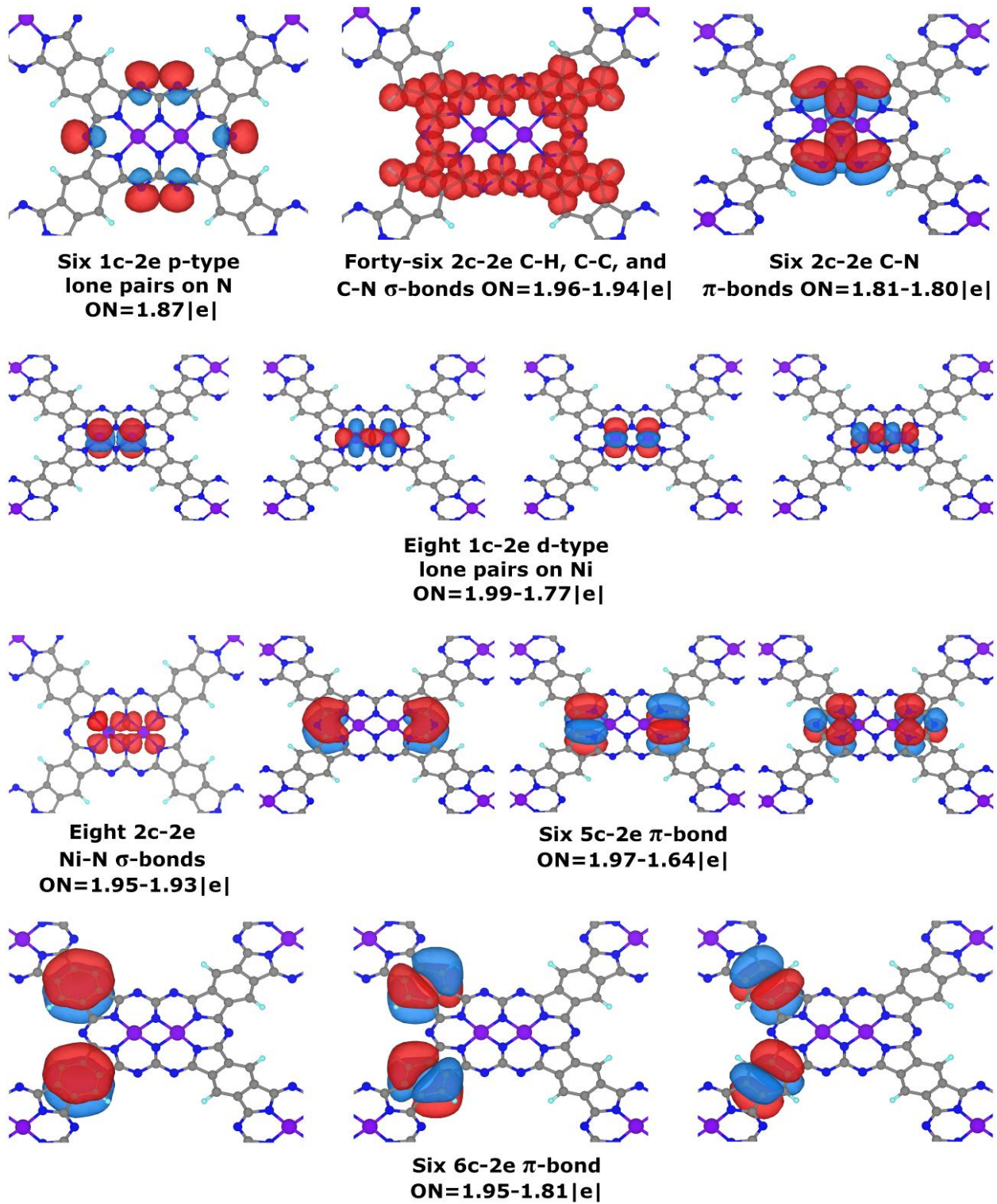

**Figure 2.** Chemical bonding pattern of the 2D Ni$_2$EPc monolayer from the SSAdNDP analysis.

### 3.4. Magnetic properties

*3.4.1. Intrinsic magnetism without biaxial strain*

In the magnetic calculations, we considered different magnetic coupling states and obtained four linearly independent magnetic configurations: FM, AFM1, AFM2, and AFM3 states, as shown in Figure 3a–d. By comparing the relative energies of FM, AFM1, AFM2, and AFM3 states, we found that the magnetic ground states for $V_2$– and $Cr_2$–EPc systems are AFM1 and FM configurations, respectively, whereas the other magnetic systems, namely, $Ti_2$– and $Mn_2$– to $Co_2$–EPc, prefer to AFM3 magnetic ordering. According to the magnitudes of average magnetic moments accumulated on each metal atom, these magnetic systems can be divided into three categories. For $Ti_2$EPc system, it has a small magnetic moment of 0.95 $\mu_B$; while for $V_2$– and $Co_2$–EPc systems, they are moderate magnetism with magnetic moments of 2.00 and 1.15 $\mu_B$. For $Cr_2$–, $Mn_2$–, and $Fe_2$–EPc systems, they have high magnetic moments of 3.29, 3.55, and 2.49 $\mu_B$, respectively.

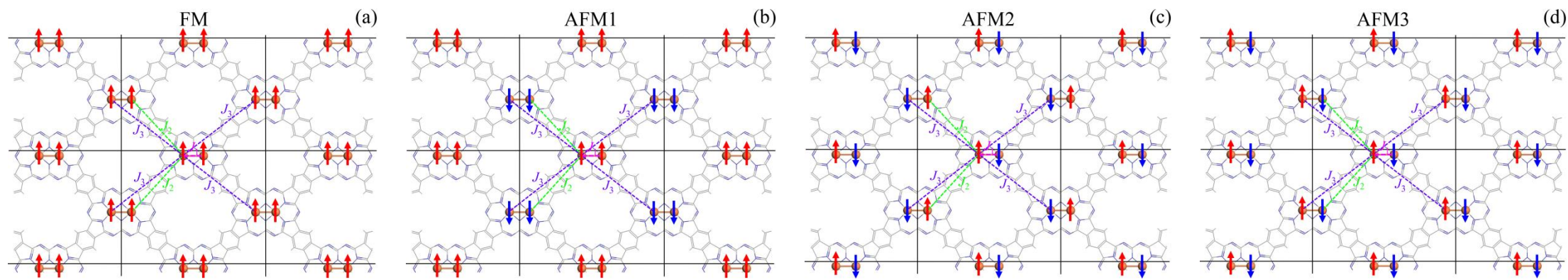

**Figure 3.** Schematic diagrams for different magnetic ordering configurations in 2D TM$_2$EPc monolayers. (a) ferromagnetic (FM), (b) antiferromagnetic-1 (AFM1), (c) antiferromagnetic-2 (AFM2), and (d) antiferromagnetic-3 (AFM3). Orange balls are TM atoms, and arrows indicate spin directions of the TM atoms. The nearest ($J_1$), next-nearest ($J_2$) and third-nearest ($J_3$) exchange interactions are indicated by magenta, green and purple dotted lines, respectively.

To explore the stability of magnetic ordering at elevated temperatures, we calculate the exchange-interaction parameters by comparing the energies of different AFM configurations with a reference to the FM configuration and use these exchange parameters to predict the Curie or Néel temperatures ($T_C/T_N$). Herein, we adopted the classical spin Hamiltonian to fit the energies of different magnetic configurations. Considering the nearest, next-nearest, and third-nearest magnetic exchange interactions, the Hamiltonian can be written as

$$H = -\sum_{i,j} J_1 M_i \cdot M_j - \sum_{k,l} J_2 M_k \cdot M_l - \sum_{m,n} J_3 M_m \cdot M_n \tag{1}$$

where $J_1$, $J_2$, and $J_3$ are the nearest, next-nearest, and third-nearest exchange parameters, $M_i$ is the magnetic moment at sites $i$, $(i, j)$, $(k, l)$, and $(m, n)$ are nearest, next-nearest, and third-nearest sit pairs, respectively. Based on this Hamiltonian, the magnetic energies for different magnetic orderings can be evaluated by eqs (1)–(4). Then with the exchange energies defined as AFM – FM energy difference calculated by VASP, the exchange parameters $J_1$, $J_2$, and $J_3$ are obtained according to eqs (5)–(7). The calculated exchange energies and parameters are listed in Table S6. Positive (negative) exchange energy indicates that the ground state of the system is FM (AFM). As addressed in Table S6, our results indicate that the TM atoms favor AFM3 coupling for $Ti_2$–, $Mn_2$–, $Fe_2$–, and $Co_2$–EPc systems, and the energies of AFM3 states are 56.56, 891.11, 675.53, and 50.13 meV lower than that of corresponding FM states, respectively. For $V_2$EPc monolayer, AFM1 state is more favorable, while in $Cr_2$EPc monolayer, the Cr atoms prefer to FM ordering.

$$E(\text{FM}) = E_0 - 2J_1 M^2 - 4J_2 M^2 - 8J_3 M^2 \tag{2}$$

$$E(\text{AFM1}) = E_0 + 2J_1 M^2 - 4J_2 M^2 + 8J_3 M^2 \tag{3}$$

$$E(\text{AFM2}) = E_0 + 2J_1 M^2 + 4J_2 M^2 - 8J_3 M^2 \tag{4}$$

$$E(\text{AFM3}) = E_0 - 2J_1 M^2 + 4J_2 M^2 + 8J_3 M^2 \tag{5}$$

$$J_1 = \frac{[E(\text{AFM1}) - E(\text{FM})] + [E(\text{AFM2}) - E(\text{FM})] - [E(\text{AFM3}) - E(\text{FM})]}{8M^2} \tag{6}$$

$$J_2 = \frac{[E(\text{AFM2}) - E(\text{FM})] + [E(\text{AFM3}) - E(\text{FM})] - [E(\text{AFM1}) - E(\text{FM})]}{16M^2} \tag{7}$$

$$J_3 = \frac{[E(\text{AFM1}) - E(\text{FM})] + [E(\text{AFM3}) - E(\text{FM})] - [E(\text{AFM2}) - E(\text{FM})]}{32M^2} \tag{8}$$

Where $E_0$ is the reference energy without magnetic order. To develop a practicable spintronic device, the magnetic transition temperatures of materials should be comparable to or higher than room temperature. Therefore, we computed Curie or Néel temperatures ($T_C/T_N$) for 2D $TM_2$EPc monolayers by employing Monte Carlo (MC) simulations. Apart from 2D $Cr_2$–, $Mn_2$–, and $Fe_2$–EPc monolayers, the magnetic transition temperatures of other systems are all lower than room temperature (see Table S6). The temperature-dependent specific heat ($C_v$) obtained from MC simulations is shown in Figure 4. It is obvious that the $T_C$ of Cr$_2$EPc monolayer is about 221 K, and the $T_N$ of Mn$_2$– and Fe$_2$–EPc monolayers are around 217 and 325 K, respectively. The magnetic transition temperatures of Cr$_2$–, Mn$_2$–, and Fe$_2$–EPc monolayers are higher than previously reported 2D metal-organic sheets containing one isolate metal center, shuch as MnPc (~150 K),[9] CrPc (140 K),[44] FePc (130 K),[44] and Cr–porphyrin (187 K).[45] High magnetic transition temperatures render these novel TM$_2$EPc sheets very attractive for future applications in low-dimensional spintronics.

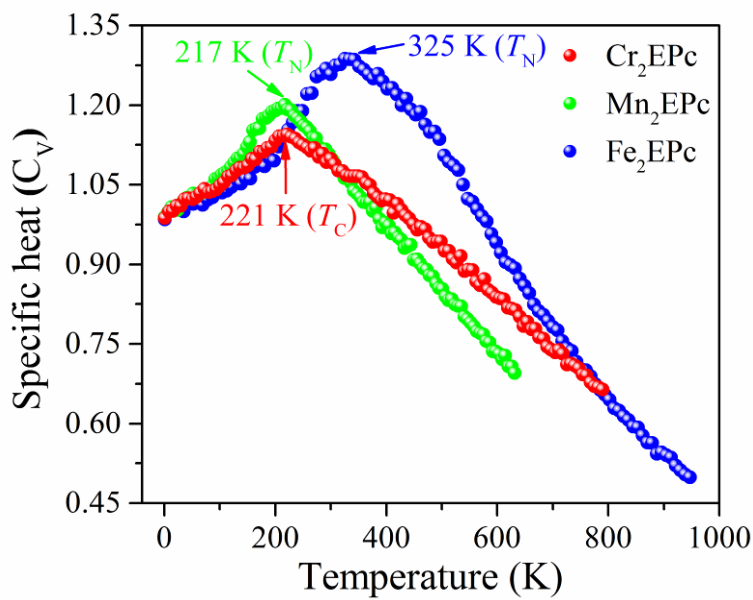

**Figure 4.** Evolution of specific heat ($C_v$) with respect to temperature for the 2D Cr$_2$–, Mn$_2$–, and Fe$_2$–EPc monolayers from Monte Carlo (MC) simulations. The corresponding magnetic transition temperatures ($T_C/T_N$) are also displayed.

In order to understand the origination of the magnetism, we analyzed the spatial distribution of spin densities ($\Delta\rho = \rho_\uparrow - \rho_\downarrow$), as present in Figure 5 and S20. It is found that the magnetic moment mainly stems from the central TM atoms, and the neighboring N and C atoms on the phthalocyanine rings are also slightly spin-polarized, contributing a little magnetic moment. For example, in the ferromagnetic Cr$_2$EPc system, all the Cr atoms have the same spin orientation with a large magnetic moment of 3.29 $\mu_B$, and the neighboring N atoms are spin-polarized in opposite orientation with a weak magnetic moment of –0.28 $\mu_B$. The C atoms which are coordinated with the N are slightly spin-polarized in opposite orientation with a very small magnetic moment of 0.02 $\mu_B$, and other N, C, and H atoms are barely spin-polarized (see Figure 5a). In the antiferromagnetic Mn$_2$– and Fe$_2$–EPc systems (Figure 5b and 5c), the magnetic moments of adjacent TM atoms are coupled oppositely. It is noteworthy that the N atoms connected to one TM atom have the opposite spin densities to that of the metal atom, while the N atoms connected to both TM atoms have spin-up and spin-down densities simultaneously. The neighboring C atoms are hardly spin-polarized. This indicates that the spin orientation of one atom is closely related to the magnetic ordering of adjacent atoms and its magnetic moment is significantly influenced by the local magnetic coupling environment. Similar phenomena are also observed in $Ti_2$–, $V_2$– and $Co_2$–EPc systems, as shown in Figure S20.

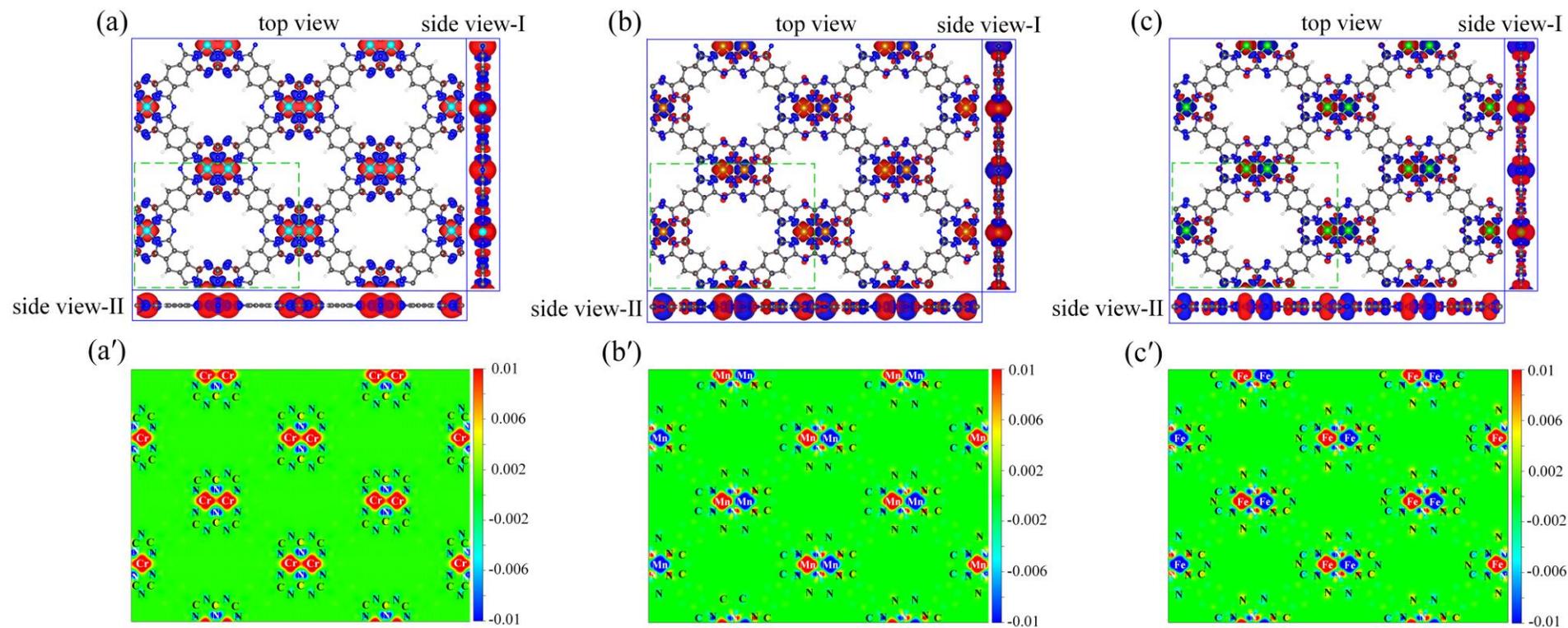

**Figure 5.** (a)–(c) Isosurfaces at the value of 0.001 e/Å$^3$ of spin densities ($\Delta\rho = \rho_\uparrow - \rho_\downarrow$) for 2D periodic Cr$_2$–, Mn$_2$– and Fe$_2$–EPc monolayers in 2 × 2 × 1 supercells; (a′)–(c′) are corresponding 2D slices of the top views in (a)–(c). Red and blue bubbles represent spin-up and spin-down densities, respectively.

According to the crystal field theory, in a crystal field environment with a square or square pyramidal symmetry, the 3d orbitals of the TM atoms split into $d_{z^2}$, $d_{xy}$, and $d_{x^2-y^2}$ orbitals and a doubly degenerate orbital ($d_{xz}$, $d_{yz}$). For the TM atoms in the $TM_2EPc$ monolayers, the absence of the square symmetry leads to five singly degenerate orbitals, as addressed in Figure 6 and S21. Based on the number of alpha and beta electrons of TM atoms in the above chemical bonding analysis results, the magnetic moment of the $TM_2EPc$ monolayers can be understood based on the simple "4 + 1" splitting model. In such a splitting model, the d orbital is approximately treated into low-lying quadruple-degenerate g1 orbitals ($d_{z^2}$, $d_{xz}$, $d_{yz}$, and $d_{x^2-y^2}$) close in energy and one high-lying g2 orbital ($d_{xy}$). The g2 orbital can be occupied once the g1 orbitals are filled. Such a d orbital filling mode leads to magnetic moments of 0 $\mu_B$ (Sc), 1 $\mu_B$ (Ti), 2 $\mu_B$ (V), 3 $\mu_B$ (Cr), 3 $\mu_B$ (Mn), 2 $\mu_B$ (Fe), 1 $\mu_B$ (Co), 0 $\mu_B$ (Ni), 0 $\mu_B$ (Cu), and 0 $\mu_B$ (Zn) per TM atom for these $TM_2EPc$ monolayers, which are consistent with the DFT calculated results. To further unveil the underlying mechanism determining magnetic ground states and fundamentally understand the intrinsic magnetic coupling interactions from atomic level, we perform in-deep analyses of electronic/magnetic structures and chemical environment of four representative materials ($Cr_2-$, $Fe_2-$, $V_2-$, and $Mn_2-EPc$) from local coordination environment (LCE), projected density of states (PDOS), crystal/exchange field splitting (CFS/EFS), d orbital filling (DOF) and their exchange coupling interactions (Figure 6 and S21). Due to limited space, we are unable to discuss all cases in the main text, but focus on two representative examples, i.e., ferromagnetic $Cr_2EPc$ and antiferromagnetic $Fe_2EPc$ with the highest $T_C$ (221 K) and $T_N$ (325 K) in all predicted $TM_2EPc$ monolayers, respectively.

Let's first discuss the ferromagnetic $Cr_2EPc$ monolayer. From Figure 6a, we can see that the $Cr^{3+}$ is in the LCE of tetragonal planar crystal field. The PDOS analysis indicate that the d orbitals are split into five singly degenerate orbitals (Figure 6b). Among the five orbitals, $d_{z^2}$, $d_{yz}$, and $d_{x^2-y^2}$ are half occupied, leaving $d_{xz}$ and $d_{xy}$ empty, thus the electron configuration of $Cr^{3+}$ is $(d_{z^2})^1(d_{yz})^1(d_{x^2-y^2})^1(d_{xz})^0(d_{xy})^0$, as shown in Figure 6c. This gives rise to a local magnetic moment of 3 $\mu_B$, in good agreement with the result of DFT calculations (3.29 $\mu_B$). The different hierarchies of exchange interactions $J_i$ ($i = 1 \sim 3$) are labeled in Figure 6d-6f. For the nearest neighbor interaction $J_1$, the Cr-Cr distance is 2.55 Å. However, due to the Pauli repulsion effect, there is no direct exchange between the two Cr atoms with the same electron spin (electrons cannot hop between the two Cr sites), and only superexchange coupling can be carried out through the N atom as an intermediate medium. The electron configuration of the medium N atom being $p^4$ with two half filled p orbitals involved in the superexchange. Due to the Cr-N-Cr angle (84.69°) being close to 90° and short Cr-Cr distance of 2.55 Å in the superexchange path shown in Figure 6g, this type of superexchange interaction is expected to be strong FM. For next-nearest neighbor interaction $J_2$, the distance between the two Cr atoms is 10.66 Å. The large Cr-Cr distance indicates the magnetic interaction of the two Fe atoms is mediated through the p electron of EPc moiety (Figure 6h). The p-d hybridization between Cr and EPc moiety as well as the π conjugation in the EPc framework are beneficial to form stable FM order. The very large third-nearest neighbor Cr-Cr distance (12.57 Å) leads to weak Cr-EPc-Cr superexchange interaction, which produces rather weak FM coupling (Figure 6i). Three types of superexchange interactions collectively determine the ferromagnetic ground state of the $Cr_2EPc$ monolayer and results in high $T_C$.

For the situation of antiferromagnetic $Fe_2EPc$, the LCF of magnetic Fe atoms (Figure 6j) and d orbital splitting (Figure 6k) are similar to the situation of Cr atoms in ferromagnetic $Cr_2EPc$ monolayer. However, due to the significantly different electronic configuration of $Fe^{2+}$ being $(d_{yz})^2(d_{xz})^2(d_{x^2-y^2})^1(d_{z^2})^1(d_{xy})^0$ (Figure 6l), these magnetic exchange interactions lead to an AFM ground state in the $Fe_2EPc$ monolayer. The AFM ground state can be understood from the competition among different types of exchange interactions (direct exchange and superexchange, Figure 6m-6o). The direct exchange ($J_1$) come from the direct electron hopping of two half occupied d orbitals with opposite electron spin (Figure 6p). This leads to strong AFM coupling. In the next-nearest neighbor interaction ($J_2$), the Fe-Fe distance is 10.58 Å, which is too far to couple directly between the two Fe atoms. It should be the superexchange interaction with one full filled p orbital of EPc moiety as intermediate (Figure 6q). Such d-p-d superexchange interaction is expected to be rather strong AFM. The magnetic coupling interaction between the two third-nearest neighbor Fe atoms ($J_3$) is mediated through one half occupied p orbital of EPc moiety (Figure 6r) with very large Fe-Fe distance of 12.48 Å, which results in weak FM. The net effects of these interactions (direct exchange and superexchange) lead to the antiferromagnetic ground states of $Fe_2EPc$ monolayer with high $T_N$.

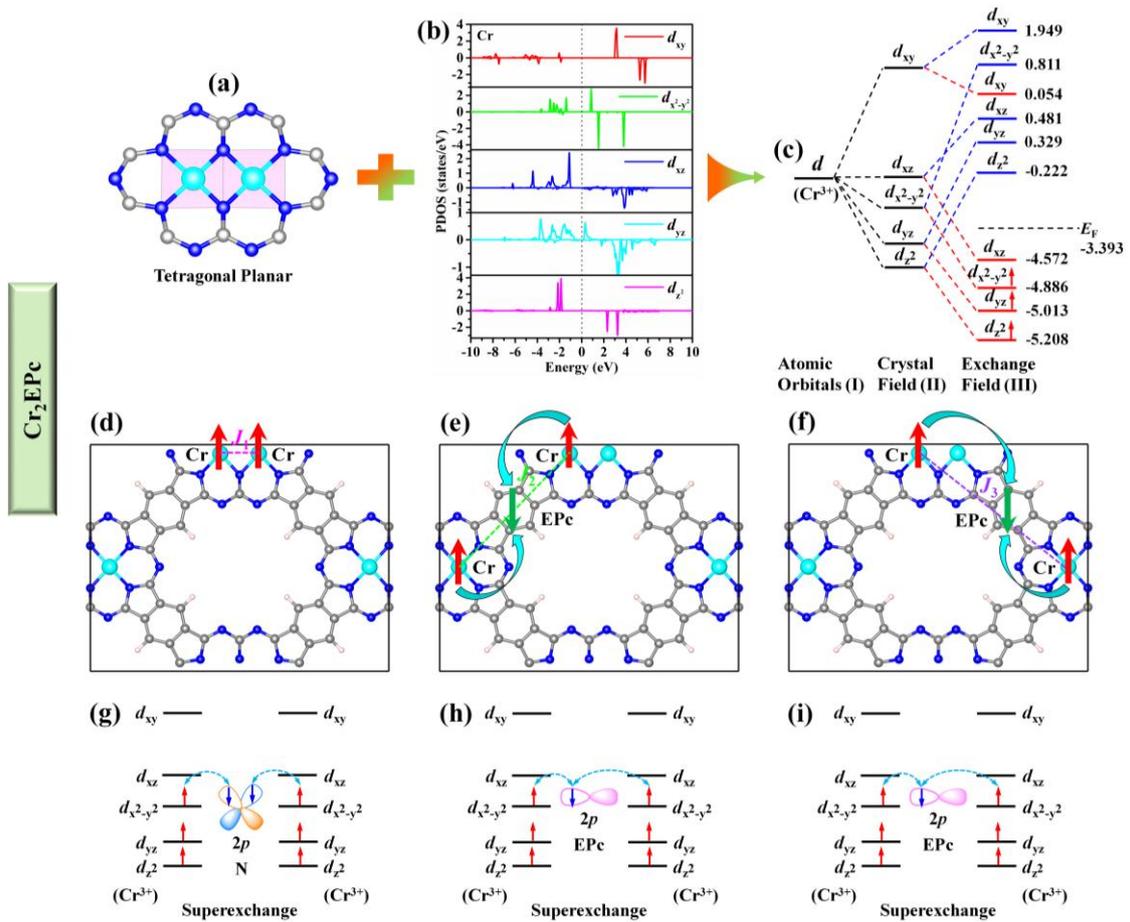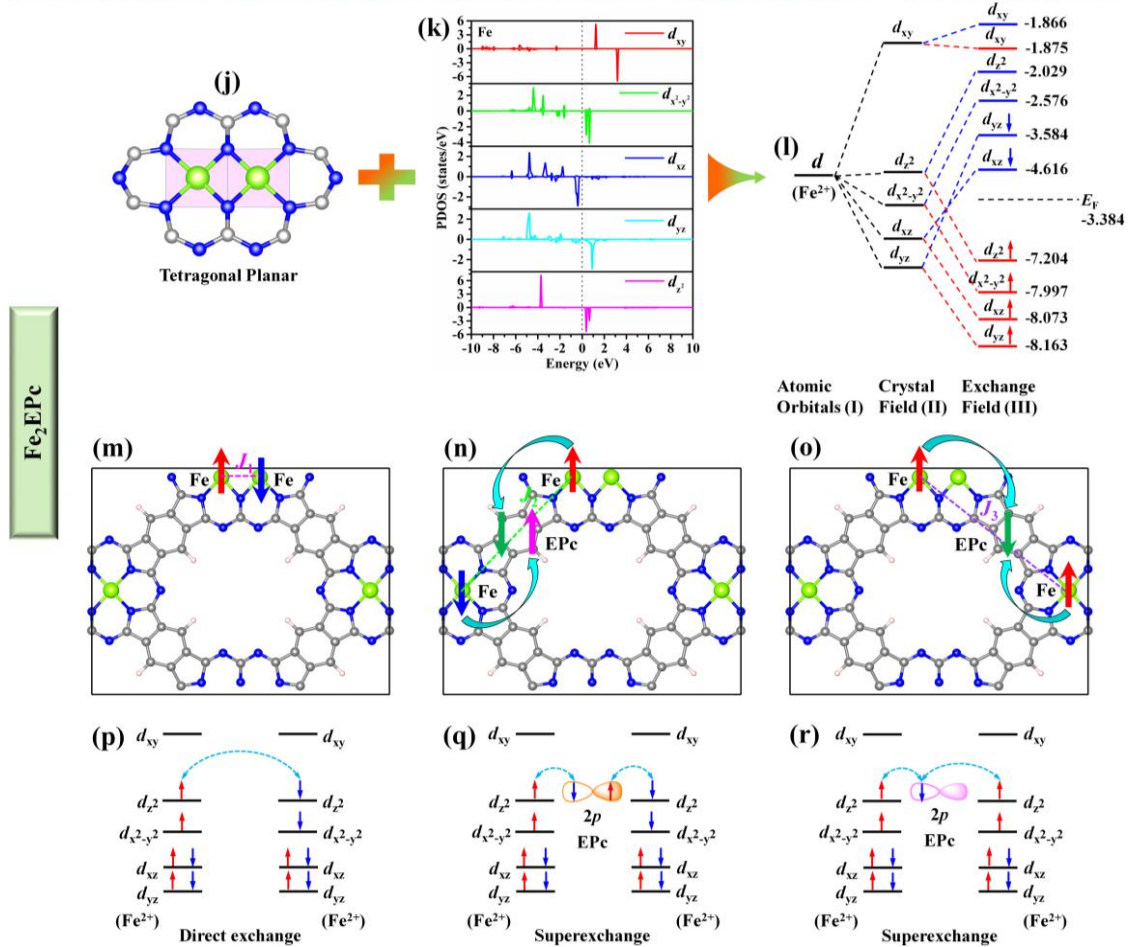

**Figure 6.** The underlying mechanism of magnetic coupling in two representative materials $Cr_2EPc$ and $Fe_2EPc$ monolayers. (a)/(j) are the local coordination environment of magnetic atoms, (b)/(k) are the PDOS as electronic fingerprint used to distinguish the degeneracy of five d orbitals of magnetic atoms, (c)/(l) are the splitting of five d orbitals of magnetic atoms in both crystal and exchange fields, (d)-(f)/(m)-(o) are the schematic diagrams of magnetic coupling and its relevant atoms, (g)-(i)/(p)-(r) are the distribution of d electrons in different d orbitals and the exchange coupling among the electrons located in different orbitals.

*3.4.2. Effect of biaxial strain on magnetism*

The above discussion and analysis are focus on the intrinsic magnetism at equilibrium structure without external stimuli. In this part, we will explore the modulation of magnetism (with possible magnetic phase transition and substantial improvement of critical temperature) through elastic strain engineering as it is one of the most effective approaches to tune the magnetism and critical temperature of 2D materials, which has been confirmed in many reports in literature.[46-49] Thus, we have confidence that the elastic strain engineering can bring unexpected surprises for the $TM_2EPc$ monolayers. In this study, in-plane biaxial strains from −10% to 10% were applied to $TM_2EPc$ monolayers to explore the regulation effects on the magnetic coupling and critical temperature. Since $Sc_2$–, $Ni_2$–, $Cu_2$–, and $Zn_2$–EPc remain nonmagnetic state under strain, we focus on the remaining six magnetic systems. Interestingly, some materials exhibit multiple magnetic phase transitions within the strain range of −10% ~ 10% in Figure S22. Here, four representative examples of $V_2$–, $Cr_2$–, $Mn_2$–, and $Fe_2$–EPc are discussed in detail. One can see that the magnetic moments of all systems increase monotonically from −10% to 10% strains in Figure S22a–S22d. This surprising phenomenon can be perfectly explained by the spin density plots at different strains in Figure 7. Among the four representative examples, three materials (except $Mn_2EPc$) undergo the magnetic phase transition at different thresholds (−4%/−6%/4%/6% for $V_2EPc$, −8%/4% for $Cr_2EPc$, −6% for $Fe_2EPc$). The magnetic transition temperatures of all four materials could be improved from 90, 221, 217, and 325 K to 150, 270, 295, and 350 K at −10%, −4%, −4%, and −6% strains for $V_2$–, $Cr_2$–, $Mn_2$–, and $Fe_2$–EPc, respectively. The magnetic ground states of $Mn_2$–, $Ti_2$–, and $Co_2$–EPc monolayers are retained AFM state over the entire strain range (Figure S22e–S22f), indicating that they are insensitive to external strain, which paves the way towards anti-disturbant antiferromagnetic spintronics.

The above discussion and analysis demonstrate that elastic strain engineering is a very powerful way to substantially modulate the magnetism of $TM_2EPc$ monolayers. This laies the foundation towards diverse applications (e.g., spin filter, spin field effect transistor, magnetic tunnel junction, magnetic memory resistor, magnetic storage, magnetic field sensor, magnetic random-access memory, etc) of these magnetic materials in various occasions.[50-53]

*3.4.3. The microscopic mechanism of modulation of magnetism via biaxial strain*

The above exciting results demonstrate that elastic strain engineering is a very powerful routine towards playing around the magnetism of $TM_2EPc$ monolayers with desirable magnetic properties and functionalities in (potential) diverse applications. It is quite crucial to unveil the origin from the molecular level, which will provide fundamental understanding on the microscopic mechanism of magnetism and shed guiding insights on the practical applications of $TM_2EPc$ monolayers implemented spintronic devices under different circumstances. Here, we propose two simple and effective descriptors to unveil the origin of modulation of magnetism.

After many efforts and attempts of trial and errors, we unexpectedly uncover that the plots of spin density (Figure 7) can be very clearly used to distinguish the different magnetic states and the magnitude of magnetic moments, and effectively elucidate the evolution of the rich magnetic phenomena caused by elastic strain engineering. From Figure 7 one can clearly see the magnetic phase transition at specific stain and the vivid evolution of spin density (magnetic moment). These plots can quite well explain the fact that the magnetic moments (Figure S22) increase monotonically from −10% to 10% strains. This is can be explained phenomenologically through the hypothesis that the compressive strain will lead to the decrease of the distances between adjacent atoms, which results in the increase of the overlap between atomic orbitals. The increase of atomic orbital overlap benefits to the electron pairing, which will lead to the increase of the number of paired electrons, thus, the number of single electrons decreases, i.e., the decrease of magnetic moment. The process of gradually increased red and blue bubbles in each panel in Figure 7 indicate the evolution trend of spin density (magnetic moment) from small to large, and vice versa. In short, the spin density plot can be used as a novel, simple, effective and unified "graph descriptor" to intuitionistically reflect the origin of regulation of magnetism and magnetic phase transition at the molecular level. This will greatly facilitate the study of magnetism, such as, quickly identify the magnetic ground state and magnetic phase transition, intuitively explain the evolution of magnetic moment under strain, and fundamentally deepen the understanding of the nature of magnetic exchange interactions under strain.

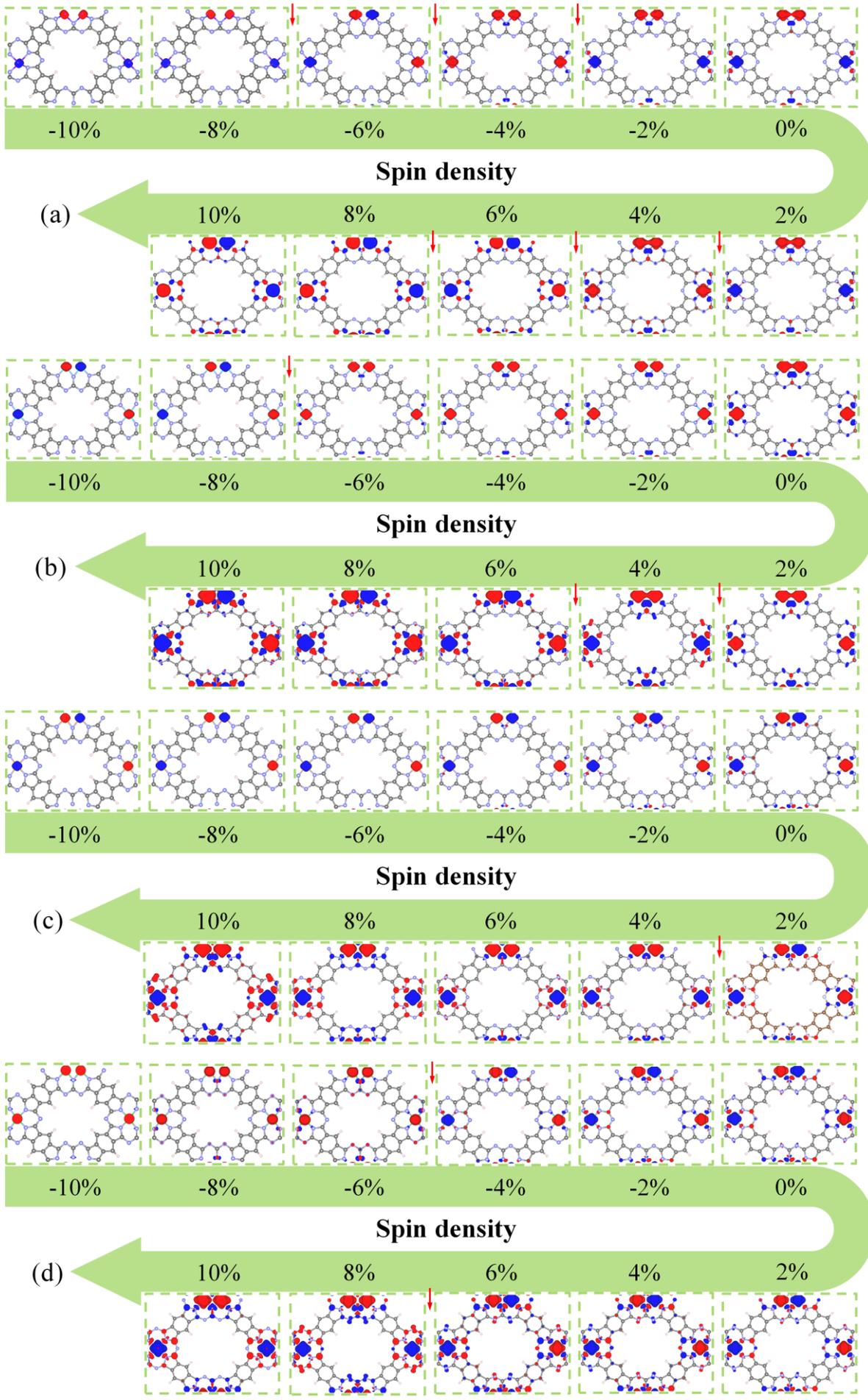

**Figure 7.** The plots of spin density (around the large green arrow) of four representative materials at different strains from −10% to 10% with interval of 2%. (a) V$_2$EPc, (b) Cr$_2$EPc, (c) Mn$_2$EPc, (d) Fe$_2$EPc. The red arrows indicate the magnetic phase transition.

From charge distribution analysis, the Bader charge (Figure S23) located on the TM atom feature monotonically increasing from −10% to 10% stains, whereas the Bader charge on EPc monolayer display monotonically decreasing from −10% to 10% strains. This indicates that the electrons flow from TM atom to EPc monolayer under a series of strains. The larger the strain (from −10% to 10%), the more electron lose. The electron flow among different atoms demonstrates the change of electron filling in different orbitals, which further indicates the evolution of electron occupation number (EON) in different orbitals. In TM$_2$EPc monolayers, N, C, and H are the main group elements, the magnetism of TM$_2$EPc monolayers is mainly contributed by the 3d orbitals of TM atoms with negligible contribution from N and C atoms. So, we focus on exploring the evolution of EON of five d orbitals of TM atoms. The calculated results of EON of d orbitals at different critical strains are collected in Figure S24. From there one can clearly see the evolution of EON of spin up and down channels of five d orbitals. Interestingly, the EON of one spin channel increases, the other spin channel decreases for the same d orbital, and vice versa. The larger the difference of EON between two spin channels, the larger the magnetic moment. The crosspoint of two spin channels reflects the possible magnetic phase transition, the number of crosspoint (Figure S24) implies the number of magnetic phase transitions, which further indicates the number of different magnetic states. Let's take V$_2$EPc as an example for discussion. The spin plots (Figure 7a) indicate 6 times magnetic phase transitions associated with seven magnetic states. Surprisingly, this is consistent well with the EON plot in Figure S24a. From Figure S24a, one can clearly see the strain intervals where magnetic phase transition occurs, they are <−8%, −6%>, <−6%, −4%>, <−4%, −2%>, <2%, 4%>, <4%, 6%>, and <6%, 8%>. The similar situation can be seen for other TM$_2$EPc monolayers, indicating the general applicability. Thus, EON can be regarded as a new, simple, and effective "data descriptor", which will be very convenient to reveal the electron filling in d orbitals, the magnetic exchange mechanism, spin polarizability, the mutual evolution of charge and spin, and effectively correlate the two degrees of freedom (i.e., spin and charge).

### 3.5. Electronic properties
*3.5.1. Intrinsic electronic properties without biaxial strain modulation*

Having identified the intrinsic stability and chemical bonding pattern of the 2D TM$_2$EPc monolayers, we then investigated their electronic performance by computing the band structures and corresponding total and projected density of states (TDOS and PDOS), and found a series of very interesting electronic phenomena, including half-metallicity, Dirac electronic state, and semiconducting properties.

**Half-metallicity:** Half-metals,[54] with one spin channel conducting and the other semiconducting, have been consider as the ideal materials for spintronic applications. To develop practicable spintronic devices with half-metals, besides the Curie temperature should be comparable to or higher than room temperature, another essential issue is that the half-metallic energy gap should be wide enough (e.g., >1.0 eV) to efficiently prevent the spin-flip transition of carriers due to thermal excitation and preserve half-metallicity in the working temperature window.[55] By spin-polarized band structure calculations, we surprisingly found that the $Cr_2EPc$ monolayer exhibits interesting half-metallic properties. From Figure 8a, we can see that the spin-up state is semiconducting, while the spin-down state is metallic, with two bands (valence and conduction bands) crossing the Fermi level. At PBE+U level, the half-metallic energy gap is 0.878 eV. In order to get reliable band structures and accurate band gap values, the hybrid functional which includes exact screened HF exchange interaction, was employed for more accurate electronic properties calculations. The HSE06 functional results show that the $Cr_2EPc$ monolayer is still a half-metal with a considerably direct spin-up energy gap of 1.416 eV. The half-metallic energy gap is larger than some reported half-metals, such as $Sr_2FeMoO_6$ (~0.8 eV)[56], $Mn_2FeReO_6$ (~1.0 eV),[57] $LaCu_3Fe_4O_{12}$ (~0.82 eV),[58] and is comparable to the $NaCu_3Fe_2Os_2O_{12}$ (~1.6 eV).[55] Therefore, the $Cr_2EPc$ monolayer is expected to be a novel half-metal for spintronic applications.

**Dirac electronic state:** Excitingly, apart from the half-metallicity, another fascinating property, Dirac electronic state, was also found in the $Cr_2EPc$ monolayer. From the 2D band structure (Figure 8b) we can see that this Dirac cone is formed by two linear bands crossing the Fermi level located at the highly symmetric line Y–Γ, with the crossing point exactly at the Femi level, close to Γ point. For clearly illustrating the Dirac cone, we plotted the 3D band structure in Figure 8c. It is can be seen that the shape of this Dirac cone is similar to the Dirac cones of graphene. Interestingly, one Dirac cone is also observed in the metallic $V_2EPc$ monolayer (Figure 8f). From the 2D band structure (Figure 8g) we can see that this Dirac cone is located between S and Y, close to the high-symmetry point Y. It is slightly below the Fermi level. The 3D band structure (Figure 8h) shows that the Dirac cone is a distorted structure which is in sharp contrast to the Dirac cones of graphene. After performing HSE06 calculations, we found that the half-metallicity of $Cr_2EPc$ and metallic feature of $V_2EPc$ are still retained at the hybrid functional level, and their respective Dirac cone in the HSE06 band structure also remain (as shown in Figure 8e and 8j). It is noteworthy that, at HSE06 level, there are almost no changes on the Dirac cone of the $Cr_2EPc$ monolayer, but for $V_2EPc$ monolayer, the changes are considerable. The two bands crossing the Fermi level become more linear dispersion and the new Dirac cone is located between Y and Γ, with the crossing point exactly at the Femi level (see Figure 8j).

The Dirac-cone-like electronic states generally produce excellent electronic transport properties. As an important performance metric related to applications, the Fermi velocities $v_F$ of the $Cr_2$– and $V_2$–Pc monolayers were calculated according to the expression $v_F = \frac{\partial E}{\hbar \partial k}$ by using the HSE06 method, where $\frac{\partial E}{\partial k}$ is the slop of the linear band dispersion around the Dirac point and $\hbar$ is the reduced Planck's constant. At HSE06 level, the slops of the bands near the Dirac point of the $Cr_2EPc$ monolayer in $y$ direction are –2.67 and 3.15 eV Å, and corresponding Fermi velocities along the Γ → Y and Y → Γ directions are $4.06 \times 10^5$ and $4.79 \times 10^5$ m/s. For $V_2EPc$ monolayer, the slops of the bands near the Dirac point in $y$ direction are –2.10 and 3.74 eV Å, and corresponding $v_F$ values along the Γ → Y and Y → Γ directions are $3.18 \times 10^5$ and $5.68 \times 10^5$ m/s. The existence of anisotropy leads to different slops, and thus produces direction-dependent Fermi velocities. These Fermi velocities are comparable to those of silicene ($5.25 \times 10^5$ m/s),[59] germanene ($5.59 \times 10^5$ m/s),[60] and stanene ($4.40 \times 10^5$ m/s),[61] and are also in the same order of graphene ($8.22 \times 10^5$ m/s).[62] Furthermore, the linear dispersion of energy bands means that the effective mass of the carriers is very close to zero according to the definition of the effective mass of carriers $m^* = \hbar \left[ \frac{d^2 E(k)}{dk^2} \right]^{-1}$. Thus, the $Cr_2EPc$ and $V_2EPc$ monolayers are expected to be used for high-speed devices and circuits due to the ultra-high carrier mobility and massless carrier character near the Dirac point. Furthermore, the Dirac half-metallicity and intrinsic ferromagnetism feature of the $Cr_2EPc$ monolayer makes it a more fascinating material, that is, the high carrier mobility in one spin channel and a relatively large band gap in another spin channel constitute a great potential for applications in nanoelectronics and spintronics.

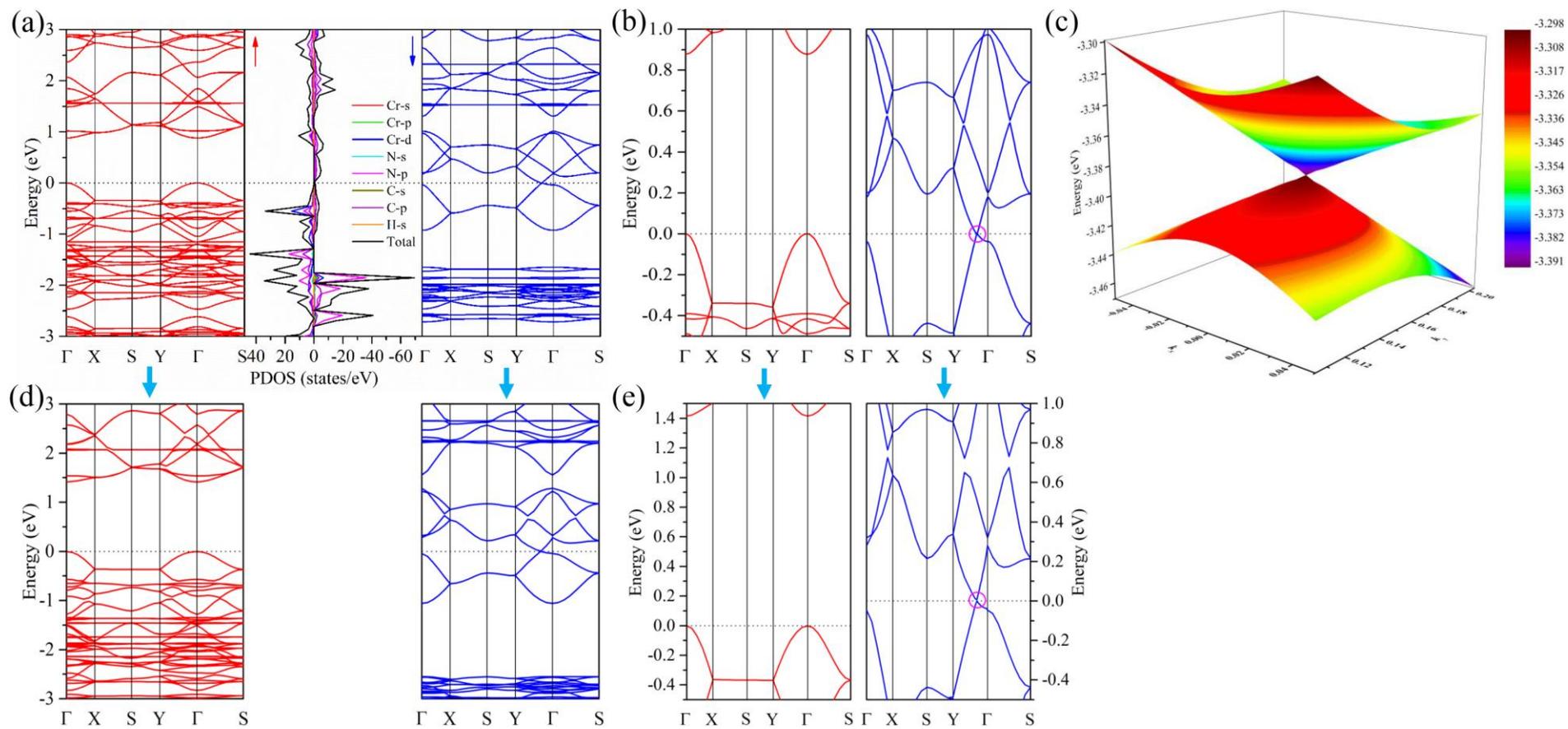

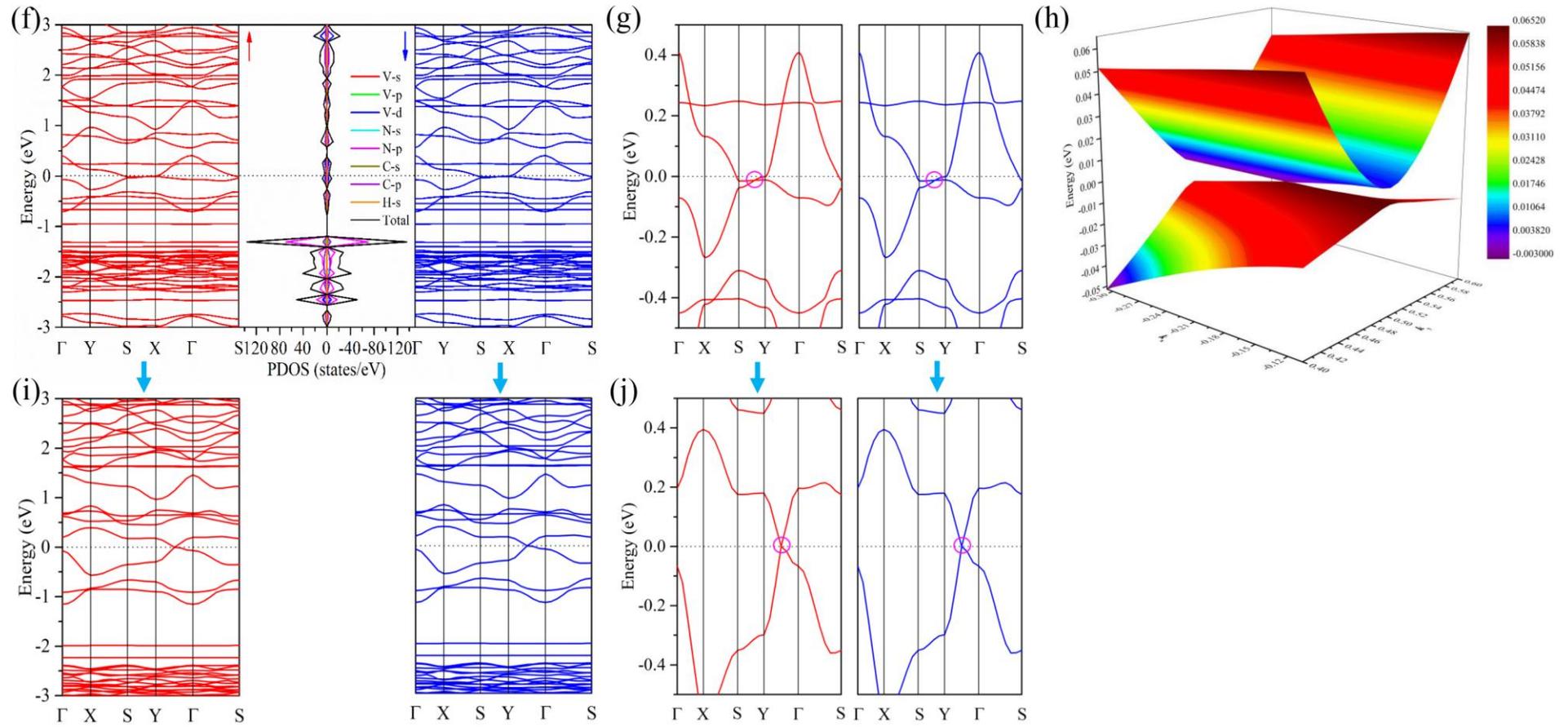

**Figure 8.** (a) and (f) are the calculated band structures and corresponding PDOS of the 2D Cr$_2$EPc and V$_2$EPc monolayers at PBE+U level. (b) and (g) are the partially zoom-in plots of the region near Fermi levels of (a) and (f). (c) and (h) are three-dimensional band structures for the demonstration of Dirac cones. (d) and (i) are the calculated band structures at HSE06 level. (e) and (j) are the partially zoom-in plots of the region near Fermi levels of (d) and (i). Dirac cones are marked with magenta circles in (c) and (h). The Fermi levels are set to zero and indicated in dark dotted lines in band structures. The red and blue arrows denote spin-up and spin-down, respectively.

**Semiconducting properties:** The band structures of other eight systems ($Sc_2$–, $Ti_2$–, and $Mn_2$– to $Zn_2$–EPc) calculated at PBE+U level are shown in Figure S25. Indirect band gaps were observed in $Sc_2$–, $Ti_2$–, and $Mn_2$– to $Ni_2$–EPc monolayers, corresponding band gap values are 0.136, 0.034, 0.241, 0.414, 0.608, and 0.932 eV, respectively. A direct band gap of 0.348 eV was found in $Cu_2$EPc monolayer. The $Zn_2$EPc monolayer is a metal. From Figure S25, we can see that the PDOS of these $TM_2$EPc monolayers are mainly contributed by N-2$p$, C-2$p$ and metals-3$d$ states, with N-2$p$ more pronounced than C-2$p$ and metals-3$d$ states, whereas the contribution of other electronic states, such as metals-3$s$, 3$p$, N-2$s$, C-2$s$, and H-1$s$ states are negligible. Furthermore, we can see that for $Sc_2$–, $Mn_2$–, $Cu_2$–, and $Zn_2$–EPc systems, the dominate contributions to valence band maximum (VBM) and conduction band minimum (CBM) are both from N-2$p$ and C-2$p$ states. For $Fe_2$–, $Co_2$–, and $Ni_2$–EPc systems, both VBM and CBM are almost exclusively contributed by N-2$p$ and TM-3$d$ states. There are apparent hybridizations between N-2$p$ and C-2$p$ as well as metals-3$d$ states, implying the existence of bonds between them, which is consistent with the chemical bonding analysis in previous discussion. To get accurate band structures of these materials, a more accurate HSE06 functional was employed to verify the semiconducting properties. The calculated results are shown in Figure S26 and Table S7. From the comparison of the band structures calculated at PBE+U and HSE06 levels, we can see that they are basically the same except the conduction bands and deep valence bands in HSE06 band structures have respectively moved to higher and lower energy regions. For $Sc_2$–, $Ti_2$–, and $Mn_2$– to $Cu_2$–EPc monolayers, the new band gaps calculated at HSE06 level are 0.225, 0.313, 0.598, 0.832, 1.114, 1.624, and 0.447 eV, respectively. While for $Zn_2$EPc monolayer, after performing hybrid functional calculations, it has become a semiconductor of 0.245 eV, which indicates that the intrinsic ground state of $Zn_2$EPc monolayer is semiconducting electronic state rather than metallic electronic state.

If taking electronic structure and magnetism into account together, the 2D $TM_2$EPc monolayers can be further divided into four categories: nonmagnetic semiconductors ($Sc_2$–, $Ni_2$–, $Cu_2$–, and $Zn_2$–EPc), antiferromagnetic metal ($V_2$EPc), ferromagnetic half-metal ($Cr_2$EPc), and antiferromagnetic semiconductors ($Ti_2$–, $Mn_2$–, $Fe_2$–, and $Co_2$–EPc).

*3.5.2. Effect of elastic strain engineering on electronic properties*

In addition to having a great influence on the magnetism and critical temperatures of the $TM_2$EPc monolayers, strain also significantly changes their electronic properties. As shown in Figures S27, we calculate the band gap evolution of each structure under biaxial strain. It can be seen that $V_2$EPc undergoes the transition process of semiconductor → metal → semiconductor during the application of −10% ~ 10% strains (Figure S27a). Specifically, $V_2$EPc is AFM metal with Dirac cone under −2% ~ 2% strains, and begins to transform into FM metal at −4% and 4% strains. When the compressive and tensile strains continue to increase to −6% and 6%, $V_2$EPc changes from FM metal to AFM semiconductor. The band gap gradually decreases with the increase of compressive strain, but monotonically increases with the increase of tensile strain. The electronic structure of $Cr_2$EPc monolayer transforms from AFM semiconductor to FM half-metal and then to AFM semiconductor under −10% ~ 10% strains (Figure S27b). $Cr_2$EPc is a half-metal with Dirac cone under −6% ~ 2% strains, and changes from Dirac half-metal electronic state to semiconductor when the compressive and tensile strains increase to −8% and 4%, respectively. Under −6% ~ 2% strains, the spin-up band gap of half-metal is gradually increased, ranging from 0.60 ~ 0.99 eV. With the tensile strain increases from 4% to 10%, the band gap of semiconductor improves from 0.26 to 0.64 eV. As shown in Figure S27c, $Mn_2$EPc presents a transition of AFM semiconductor → AFM metal → AFM semiconductor under biaxial strain. When applying −10% ~ 2% strains, $Mn_2$EPc is a semiconductor with the band gap of 0.24 ~ 0.26 eV. It becomes a metal at 4% ~ 8% tensile strains, and when the tensile strain continues to increase to 10%, it transforms into a semiconductor (band gap is 0.24 eV). $Fe_2$EPc changes from FM metal → FM half-metal → AFM semiconductor under −10% ~ 10% strains (Figure S27d). Under 4% ~ 10% strains, $Fe_2$EPc is a semiconductor, and the band gap increases first and then decreases, reaching a maximum of 0.48 eV at 8% and a minimum of 0.13 eV at 10%. When the compressive strain increases to −6%, $Fe_2$EPc transforms into half-metal. When the compressive strain increases to −10%, $Fe_2$EPc changes from half-metal to semiconductor. $Ti_2$EPc undergoes the transition of AFM metal → AFM semiconductor → AFM metal. It is metallic under compressive strain, and transforms into semiconductor under 0% ~ 6% tensile strains. When the tensile strain continues to increase to 8%, the band gap disappears and becomes metallic (Figure S27e). As shown in Figure S27f, during the entire tensile and compressive strain process, $Co_2$EPc maintains the electronic structure characteristics of the semiconductor, and the band gap gradually increases with the increase of tensile strain, reaching a maximum of 0.84 eV at 10%. Under compressive strain, the band gap increases first and then decreases, reaching a maximum of 0.63 eV at −6% (or −8%).

Rich electronic states can be obtained by applying elastic strain engineering, which indicates that strain can not only effectively modulate the magnetism, but also electronic properties. The obvious variation of spin polarization and electronic state in $TM_2$EPc monolayers under strain offers a new avenue for controllable and tunable spintronic devices.

## 3.6. Optical properties

*3.6.1. Optical absorption coefficients*

After electronic structure characterization, the 2D TM$_2$EPc can be divided into three categories: semiconductors with band gaps in the range of 0.225–1.624 eV (Sc$_2$–, Ti$_2$–, and Mn$_2$– to Zn$_2$–EPc), half metal with spin up band gap of 1.416 eV (Cr$_2$EPc), and metal with band gap of 0 eV (V$_2$EPc). To evaluate the potential performance of the 2D TM$_2$EPc monolayers (at least one spin channel has band gap except metallic V$_2$EPc monolayer) applied for optoelectronic devices such as photovoltaic cell components, we calculated the optical absorption coefficients $\alpha(\omega)$ (*xx* and *yy*) by solving the frequency dependent real and imaginary parts of the dielectric functions:

$$\alpha(\omega) = \sqrt{2}\omega[\sqrt{\varepsilon_1^2(\omega)+\varepsilon_2^2(\omega)} - \varepsilon_1(\omega)]^{1/2} \tag{9}$$

where the imaginary part $\varepsilon_1(\omega)$ was calculated by summation over empty states, and the real part $\varepsilon_2(\omega)$ was derived from the imaginary part according to the usual Kramers-Kronig transformation.[37] Finally, all components of the absorption spectra and relevant optical quantities could be determined. The obtained absorption spectra are plotted in Figure 9. The absorption coefficients of the *xx* and *yy* components are significantly different, indicating that the in-plane optical absorption has a strong anisotropy. In general, the absorption coefficient of *yy* component is larger than that of *xx* component. The absorption for Sc$_2$–, Ti$_2$–, and Cr$_2$– to Zn$_2$–EPc systems starts at 0.225 (5511), 0.313 (3962), 1.416 (876), 0.598 (2074), 0.832 (1490), 1.114 (1113), 1.624 (764), 0.447 (2774), and 0.245 eV (5061 nm), respectively, which shows that the absorption edges of all the systems appear in infrared region except for Ni$_2$Pc monolayer lying in visible light region. It can be can see that TM$_2$EPc monolayers exhibit a full response to visible light. In particular, Sc$_2$–, Ti$_2$–, Mn$_2$–, Cu$_2$–, and Zn$_2$–EPc monolayers have strong absorption in the ultraviolet and infrared regions in addition to visible light, showing extraordinary solar light-harvesting ability. In visible light region, the main absorption peaks of Sc$_2$–, Ti$_2$–, and Cr$_2$– to Zn$_2$–EPc are located at 2.80 (*yy*: 2.80), 2.50 (*yy*: 2.76), 2.48 (*yy*: 2.04), 2.98 (*yy*: 3.08), 3.07 (*yy*: 3.07), 1.65 (*yy*: 1.87), 2.65 (*yy*: 2.61), 2.79 (*yy*: 2.79) and 2.64 eV (*yy*: 2.64 eV), respectively. The absorption intensity of these materials in visible light region ranges from $1.70 \times 10^5$ to $5.32 \times 10^5$ cm$^{-1}$, especially the visible light absorption intensity of Cu$_2$EPc reaches $5.32 \times 10^5$ cm$^{-1}$. In addition, Sc$_2$–, Ti$_2$–, Mn$_2$–, Fe$_2$–, and Cu$_2$–EPc systems have high infrared absorption ability of $2.13 \times 10^5$–$3.32 \times 10^5$ cm$^{-1}$, indicating that they have strong response to infrared light and have potential prospects in infrared detectors. Overall, these 2D TM$_2$EPc monolayers have stronger absorption of ultraviolet light than infrared and visible light absorption. The ultraviolet absorption intensity of each system is above $3.0 \times 10^5$ cm$^{-1}$, and Ni$_2$EPc even exceeds $5.0 \times 10^5$ cm$^{-1}$. It is noteworthy that the absorption spectra of TM$_2$EPc monolayers vary significantly with the metal centers, thus the optical properties of the TM$_2$EPc monolayers can be tuned through substitution of diverse metal atoms, this will greatly expand the application scope in optoelectronic devices.

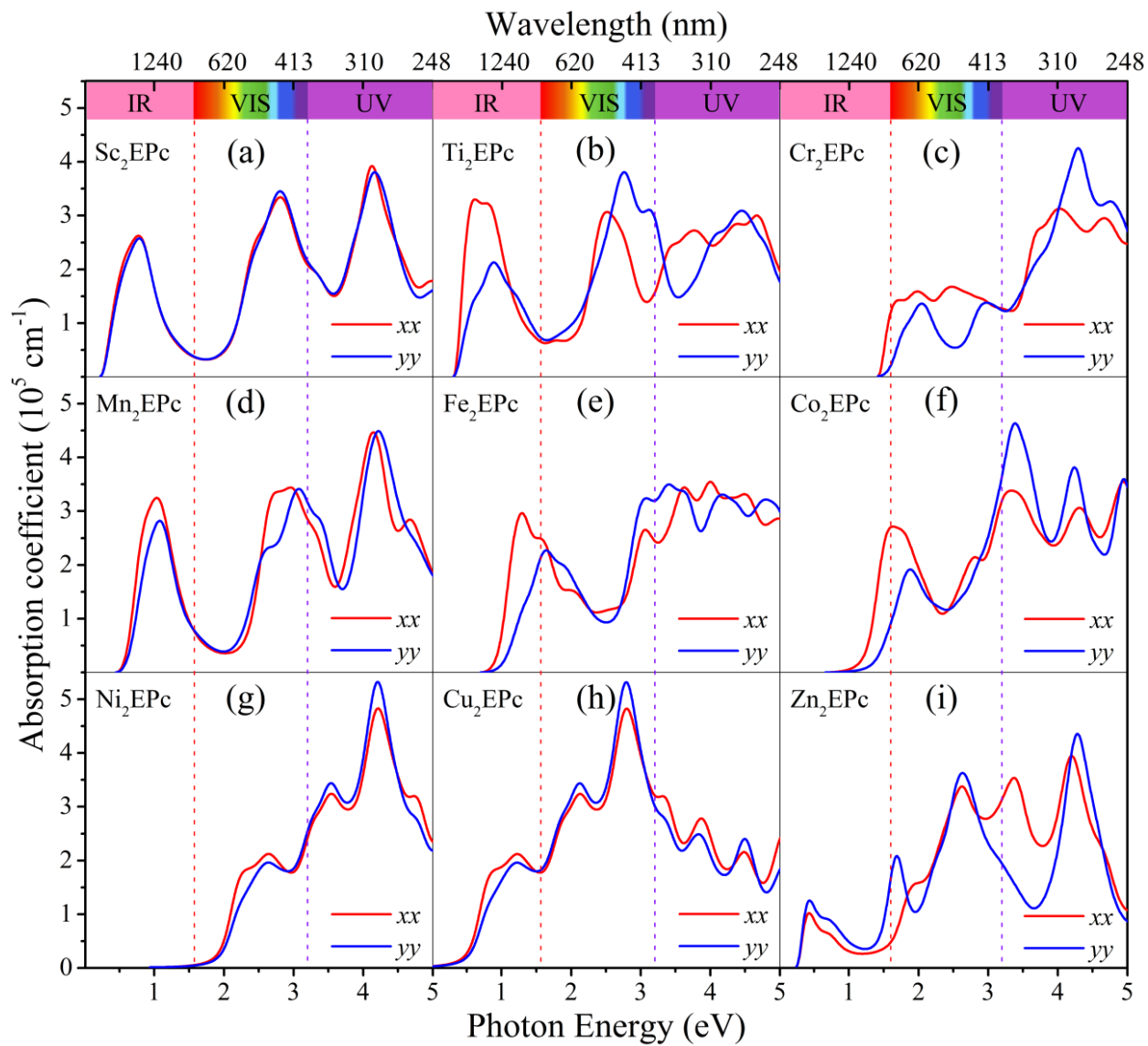

**Figure 9.** Optical absorption coefficients of 2D (a) Sc$_2$–, (b) Ti$_2$–, (c) Cr$_2$–, (d) Mn$_2$–, (e) Fe$_2$–, (f) Co$_2$–, (g) Ni$_2$–, (h) Cu$_2$–, and (i) Zn$_2$–EPc monolayers calculated at PBE+U level and corrected by HSE06 functional.

In order to study the effect of biaxial strain on optical properties in detail, we plotted the optical absorption coefficients of Fe$_2$–, Co$_2$–, and Ni$_2$–EPc monolayers under biaxial strain, as shown in Figure S34–S36. Fe$_2$EPc is a metal at −10% strain and a semiconductor with a gradually increasing band gap at −8% ~ 8% strains, so the absorption edge gradually blue-shifted (Figure S34). When the tensile strain continues to increase to 10%, the band gap begins to decreases and the absorption edge redshifts. Under 0% ~ 4% tensile strains, the absorption of visible light increases slightly, from $3.24 \times 10^5$ to $3.74 \times 10^5$ cm$^{-1}$. Under tensile and compressive strains, the absorption of Fe$_2$EPc in the ultraviolet region is significantly enhanced. The absorption intensity increased substantially from $3.57 \times 10^5$ to $5.19 \times 10^5$ and $4.27 \times 10^5$ cm$^{-1}$ at –8% and 4% strains, respectively. As depicted in Figure S35, with the increase of the amplitude of compressive strain, the band gap of Co$_2$EPc first increases and then decreases under –10% ~ –2% strains, resulting in the absorption edge first blueshifts and then redshifts. The absorption intensity of infrared light increases slightly under compressive strain, and the absorption of ultraviolet light increases dramatically from $4.65 \times 10^5$ to $7.33 \times 10^5$ at –10%. Under –2% ~ 10% strains, the band gap of Co$_2$EPc is monotonically increased, thus the absorption edge is gradually blue-shifted. At 8% strain, it has no response to infrared light, and only has absorption in the visible and ultraviolet regions. Ni$_2$EPc is a metal at –10 % strain and a semiconductor with gradually increasing band gap under –8 % ~ 10 % strains, so its absorption edge is gradually blue-shifted during the evolution of –8% ~ 10% strains (Figure S36). Under compressive strain, the absorption intensity of ultraviolet light increased sharply from $5.32 \times 10^5$ to $6.66 \times 10^5$ at −8 % strain. At the same time, with the increase of compressive strain, Ni$_2$EPc began to respond to infrared light, and the absorption of infrared light gradually increased. In addition, the absorption of visible light under compressive strain is also greatly improved, from $2.82 \times 10^5$ to $5.06 \times 10^5$ at –8 % strain. The above results show that elastic strain engineering is an effective means to manipulate the optical properties of materials. By applying biaxial strain, the absorption edge of TM$_2$EPc monolayers can be flexibly modulated, and the light absorption intensity can be significantly enhanced, especially the response to visible light and ultraviolet light, which greatly expands its application in optoelectronic and photovoltaic devices.

*3.6.2. Power conversion efficiency*

In order to evaluate the possibility of these TM$_2$EPc monolayers as components of photovoltaic devices, we constructed heterojunctions with 100 different 2D semiconductor and insulator materials (from Ref. 63 and Ref. 64). After systematic screening, only Fe$_2$–, Co$_2$–, and Ni$_2$–EPc can meet the criteria (type-II band alignment, suitable conduction band offset and energy gap) to form type-II heterojunctions with SnC, GeS, and 2H-WSe$_2$ as acceptors. The band alignment of donor and acceptor materials obtained at HSE06 level are shown in Figure 10a. To assess the performance of these heterojunctions as solar cells, we calculated their power conversion efficiency (PCE) according to formula 10.[65-66]

$$\eta = \frac{\beta_{FF} V_{oc} J_{sc}}{P_{solar}} = \frac{0.65(E_g^d - \Delta E_c - 0.3)\int_{E_g^d}^{\infty} \frac{P(\hbar\omega)}{\hbar\omega} d(\hbar\omega)}{\int_0^{\infty} P(\hbar\omega) d(\hbar\omega)} \qquad (10)$$

where $\beta_{FF}$ = 0.65 is the fill factor (FF), $J_{sc}$ is the short circuit current, and $P_{solar}$ represents the incident solar power per unit area. $V_{oc} = E_g^d - \Delta E_c - 0.3$ is the open-circuit voltage, where $E_g^d$ is the band gap of the donor material, $\Delta E_c$ is conduction band offset, which has been estimated at HSE06 level. The 0.3 eV value is an empirical parameter accounting for energy losses due to the involved kinetics in energy conversion. $P(\hbar\omega)$ is the AM1.5 solar energy flux at the photon energy $(\hbar\omega)$. As shown in Figure 10b, the calculated PCEs of the three heterojunctions Fe$_2$EPc/SnC, Co$_2$EPc/GeS, and Ni$_2$EPc/2H-WSe$_2$ are 16.25%, 19.89%, and 25.19%, respectively. In particular, the PCE of Ni$_2$EPc/2H-WSe$_2$ is higher than those of reported excitonic solar cells, such as HfSe$_2$/GeO$_2$ (22.64%),[64] BAs/BP (22.61%),[64] phosphorene/MoS2 (22.1%),[67] BP/MoS2 (20.42%),[68] GaSb/InSb (22.88%),[69] and GaAs/InAs (20.65%).[69] This indicates that Ni$_2$EPc/2H-WSe$_2$ is a promising candidate for efficient components of solar energy harvesting and photoinduced charge-carrier generation in photovoltaic cells.

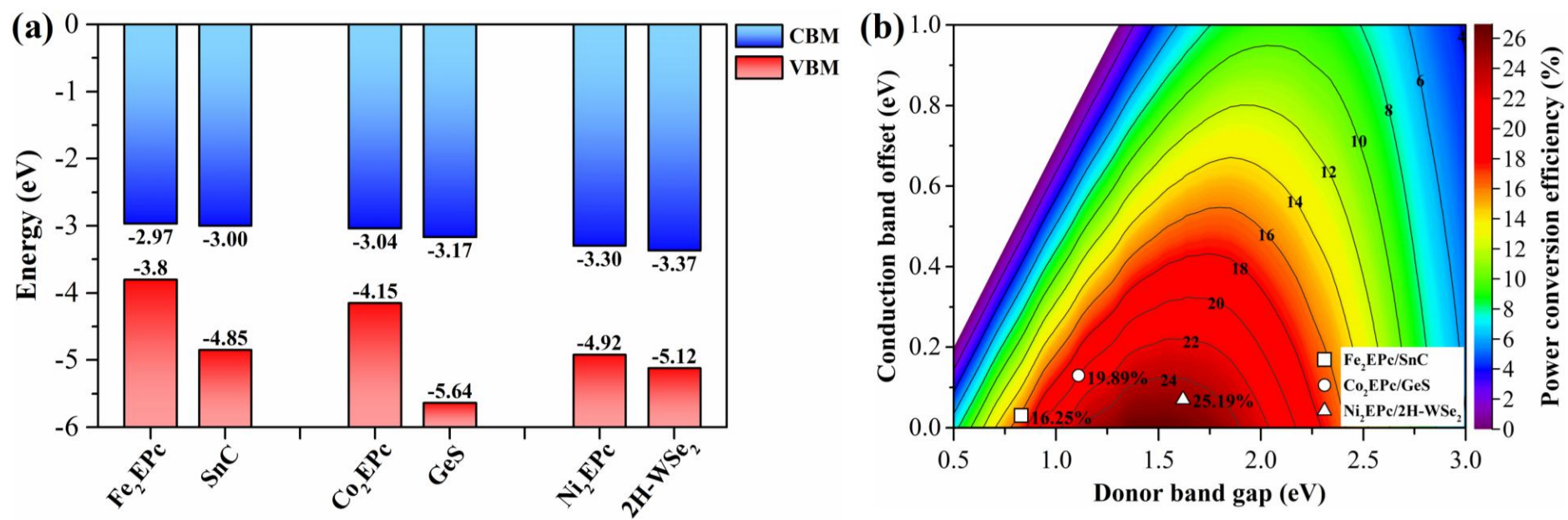

**Figure 10.** (a) Band alignment of the donor and acceptor materials obtained at HSE06 level. The values of VBM and CBM are set with respect to the absolute vacuum level. (b) Calculated power conversion efficiency (PCE) of three newly designed 2D type II heterostructures base on TM$_2$EPc monolayers. The background with different color regions refers to different efficiencies.

## 4. Conclusions

In summary, we have theoretically predicated 2D bimetal expanded phthalocyanine monolayers with high stability and fascinating properties by employing spin-polarized DFT calculations, molecular dynamics, as well as Monte Carlo simulations. Structural optimization shows that five systems, $Cr_2$–, $Mn_2$–, $Fe_2$–, $Co_2$–, and $Ni_2$–EPc, are planar configurations, whereas the remaining five systems: $Sc_2$–, $Ti_2$–, $V_2$–, $Cu_2$–, and $Zn_2$–EPc, are buckled structures. They all survived molecular dynamics simulated annealing exceeding 1200 K for 10 ps. Chemical bonding analysis shows that the TM-TM bonding is absent, and each metal center is isolated, but connected to the organic framework by four 2c-2e TM-N σ-bonds to form an extended 2D network. Electronic and magnetic calculations indicate that $Sc_2$–, $Ni_2$–, $Cu_2$–, and $Zn_2$–EPc are NM semiconductors, and $Ti_2$–, $Mn_2$–, $Fe_2$–, and $Co_2$–EPc are AFM semiconductors. Remarkably, the $V_2$EPc and $Cr_2$EPc exhibit AFM Dirac metallic and FM Dirac half-metallic features, which is quite interesting. The ultra-high Fermi velocities near Dirac cones render them promising candidates for applications in high-speed nanoelectronics and spintronics. $Cr_2$EPc can maintain the original ferromagnetic half-metallicity under –6 % ~ 2 % strains, while $Fe_2$EPc changes from antiferromagnetic semiconductor to ferromagnetic half-metal under –6 % ~ –10 % compressive strains. Excitingly, the $Cr_2$–, $Mn_2$– and $Fe_2$–EPc are predicted to have high magnetic transition temperatures of 221, 217, and 325 K, respectively, showing great potential for developing practicable spintronic devices. In addition, the intensely optical responses in visible and ultraviolet regions imply that the $TM_2$EPc monolayer have extraordinary light harvesting ability and could be highly promising applications in optoelectronic devices. The power conversion efficiency of $Ni_2$EPc/2H-$WSe_2$ is predicted as high as 25.19%, which has a great potential in photovoltaic solar cell applications. We hope that our work will stimulate the fabrication and investigation of 2D $TM_2$EPc monolayers in the near future.

**Conflicts of interest**

The authors declare no competing financial interest.

**Acknowledgements**

D.–B. L. and L.–M. Y. gratefully acknowledge support from the National Natural Science Foundation of China (21873032, 22073033, 21673087, 21903032), startup fund (2006013118 and 3004013105) from Huazhong University of Science and Technology, the Fundamental Research Funds for the Central Universities (2019kfyRCPY116, 2021yjsCXCY054), and the Innovation and Talent Recruitment Base of New Energy Chemistry and Device (B21003). The authors thank the Minnesota Supercomputing Institute (MSI) at the University of Minnesota for supercomputing resources. The work was performed on the Mesabi supercomputer at the University of Minnesota. Additional work was carried out at the Lv Liang Cloud Computing Center of China, and additional calculations were performed on the TianHe–2 supercomputer.

**Supporting Information**

Optimized ground state structures of 2D TM$_2$EPc monolayers. Phonon spectra and phonon DOS of remaining six TM$_2$EPc monolayers. AIMD simulations with the duration of 10 ps under different temperatures for 2D TM$_2$EPc monolayers. Formation energies, cohesive energies, and binding energies. Chemical bonding analysis of TM$_2$EPc molecules. Exchange energies and parameters, magnetic transition temperatures of 2D TM$_2$EPc monolayers. Spin densities of Ti$_2$–, V$_2$–, and Co$_2$–EPc monolayers in $2 \times 2 \times 1$ supercells. The evolution of electron occupation number (EON) on different d orbitals of TM atoms in four representative TM$_2$EPc monolayers. Band structures of remaining six TM$_2$EPc monolayers obtained at PBE+U and HSE06 levels. The evolution of band gaps and magnetic ground states versus biaxial strain from −10% to 10% for Ti$_2$– to Co$_2$–EPc monolayers. Calculated band structures of Ti$_2$– to Co$_2$–EPc monolayers under biaxial strain from −10% to 10%. The evolution of optical absorption coefficient versus biaxial strains for Fe$_2$–, Co$_2$–, and Ni$_2$–EPc monolayers.

"For Table of Contents Use Only"

Two-dimensional bimetal-embedded expanded phthalocyanine monolayers: a class of multifunctional materials with fascinating properties

De-Bing Long, Nikolay V. Tkachenko, Qingqing Feng, Xingxing Li, Alexander I. Boldyrev, Jinlong Yang, and Li-Ming Yang*

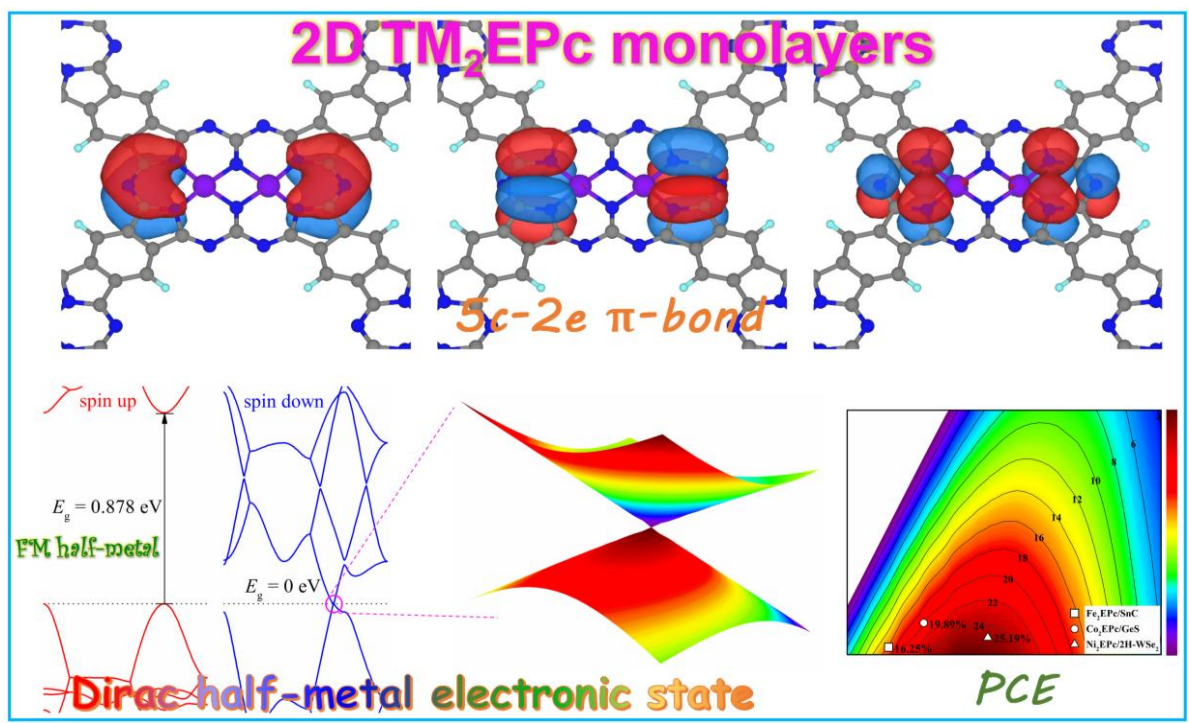

The metal-organic TM$_2$EPc monolayer can be metal, Dirac half-metal or semiconductor, and the magnetism, electronic structure and optical absorption can be tuned by embedding diverse 3d transition metal atoms in a controllable and feasible way, showing considerable flexibility and freedom in achieving versatility. The strong coordination between metal and EPc substrate accounts for the excellent structural stability. Strong visible light optical absorption and high power conversion efficiency are likely to open new doors for potential optoelectronic and photovoltaic applications.